%% file: ieee-tcomm-capacity.tex
\documentclass[journal,10pt]{IEEEtran}

\usepackage{silence}
\usepackage{premable_doc}
\usepackage{premable_tikz}
\usepackage{fancyhdr}  

\newacronym{npd}{NPD}{neural polar decoder}
\newacronym{hy}{HY}{Honda-Yamamoto}
\newacronym{bms}{BMS}{binary memoryless symmetric}

\newcommand{\nnc}[3]{\cG_\mathsf{NN}^{\left(#1, #2, #3\right)}}

\begin{document}

\title{Code Rate Optimization via Neural Polar Decoders}
\author{Ziv Aharoni,~\IEEEmembership{Member,~IEEE,}
Bashar Huleihel,~\IEEEmembership{Student Member,~IEEE,}
Henry D. Pfister,~\IEEEmembership{Senior Member,~IEEE,}
and Haim H. Permuter,~\IEEEmembership{Senior Member,~IEEE}}

\maketitle
\glsdisablehyper


\begin{abstract}
This paper proposes a method to optimize communication code rates via the application of \glspl{npd}. Employing this approach enables simultaneous optimization of code rates over input distributions while providing a practical coding scheme within the framework of polar codes. The proposed approach is designed for scenarios where the channel model is unknown, treating the channel as a black-box that produces output samples from input samples. We employ polar codes to achieve our objectives, using \glspl{npd} to estimate \gls{mi} between the channel inputs and outputs, and optimize a parametric model of the input distribution. The methodology involves a two-phase process: a training phase and an inference phase. In the training phase, two steps are repeated interchangeably. First, the estimation step estimates the \gls{mi} of the channel inputs and outputs via \glspl{npd}. Second, the improvement step optimizes the input distribution parameters to maximize the \gls{mi} estimate obtained by the \glspl{npd}. In the inference phase, the optimized model is used to construct polar codes. This involves incorporating the \gls{hy} scheme to accommodate the optimized input distributions and list decoding to enhance decoding performance. Experimental results on memoryless and \glspl{fsc} demonstrate the effectiveness of our approach, particularly in cases where the channel's capacity-achieving input distribution is non-uniform. For these cases, we show significant improvements in \gls{mi} and \glspl{ber} over the ones achieved by uniform and \gls{iid} input distributions, validating our method for block lengths up to $1024$. This scalable approach has potential applications in real-world communication systems, bridging theoretical capacity estimation and practical coding performance.
\end{abstract}


\glsresetall
\blfootnote{This paper was presented in part at the International Symposium on Information Theory (ISIT) 2024 \cite{aharoniCodeRateOptimization2024}. This research was supported in part by the National Science Foundation (NSF) under Grant 2308445 and by the NSF–Israel Binational Science Foundation (BSF) under Grant 3211/23. The views expressed are those of the authors and do not necessarily reflect those of the NSF or BSF. Source code and implementation details are available at: \url{https://github.com/zivaharoni/neural-polar-decoders}.}
\section{Introduction}

\par Channel capacity is a fundamental quantity describing the highest rate of reliable communication over a noisy channel when the transmission latency is not limited. However, knowing the channel capacity does not provide a practical coding scheme that performs well within the block lengths commonly used in contemporary communication systems. Evidently, capacity estimation and coding techniques, despite being closely related, have diverged over the years into two distinct directions: algebraic tools have been dominant in coding, while probabilistic approaches have been prevalent in capacity estimation.

\par In the context of capacity estimation, the literature is abundant with instances addressing this problem. Notable works include the Blahut-Arimoto algorithm and its extensions \cite{blahut1972computation,arimoto1972algorithm,vontobel2008generalization,naiss2012extension}, as well as algorithms for the estimation of the \gls{di} \cite{tatikonda2009capacity,permuter2008capacity,jiao2013universal,elischo2014ising}, which has been shown to be a unifying measure for the characterization of channel capacity \cite{massey1990causality}. In recent years, the rise of machine learning algorithms has significantly impacted the field of capacity estimation, providing new research directions and enhancing the capabilities of traditional algorithms. New \gls{nn}-based estimators for channel capacity have emerged, such as those in \cite{aharoni2022capacity,tsur2023neural,tsur2023data,letiziaDiscriminativeMutualInformation2022,liLearningChannelCapacity2024}. Even though there are many instances of capacity estimation techniques, none of which, except for special cases, provide practical coding schemes.

\par In this work, we aim to address this gap by devising a capacity estimation algorithm that also provides a practical coding scheme. The setting considered here assumes the absence of a channel model, treating the channel as a black-box that produces samples of channel outputs for input samples chosen by the designer. Our objectives under this setting are threefold: 1) to devise a model capable of estimating the \gls{mi} between the channel inputs and outputs from samples, 2) to optimize the input distribution with respect to the \gls{mi}, and 3) to use the optimized model to construct practical codes.

\par In order to achieve all three objectives, we employ polar codes \cite{arikan2009channel}. 
Polar codes constitute the first deterministic construction of capacity-achieving codes for \gls{bms} channels with an efficient \gls{sc} decoder. Enhancing polar codes with \gls{scl} decoding and the addition of \gls{crc} outer codes \cite{tal2015list} has dramatically improved their performance for moderate block lengths, leading to their adoption in the 5G standard \cite{3gpp2021}.
Recently, the authors proposed to use \glspl{nn} in polar codes, denoted by \glspl{npd}, in order to address cases where a channel model is not known and possibly has memory \cite{aharoni2023data_arcive}. In this work, we take one step forward and use \glspl{npd} to optimize the input distribution of the code and adapt it to list decoding.

\par The methodology proposed herein employs \glspl{npd} \cite{aharoni2023data_arcive} and uses them as a fundamental ingredient. The methodology is divided into two phases: a training phase, in which samples are used to simultaneously estimate an \gls{npd} and optimize the input distribution, and an inference phase, in which the optimized model is fixed and used to construct the code. 

\par The training phase is composed of an alternated maximization procedure, in which two core steps are repeated interchangeably. The first step is the estimation step, in which the parameters of the \glspl{npd} are updated to estimate the \gls{mi} incurred by a fixed input distribution. The second step is the improvement step, in which the \glspl{npd}'s parameters are fixed, and are used to optimize the parameters of the input distribution. This procedure continues until the convergence of the algorithm. 

\par In the inference phase, the parameters of the input distribution and the \gls{npd} are fixed and used to construct a polar code. To accommodate with the optimized input distribution, that is generally not uniform and \gls{iid}, we use the \gls{hy} scheme for the code construction, encoder and decoder \cite{honda2013polar}. Specifically, the construction of the code is performed by a \gls{mc} evaluation of the synthetic channels, and the encoder and decoder are implemented as described in \cite{honda2013polar}. To enhance the code performance, we also incorporated list decoding \cite{tal2015list} with the \glspl{npd}, which had a pivotal impact in our experiments. 

\par In the experiments, we illustrate the performance of our method on memoryless channels and \glspl{fsc}. Specifically, we showcase the estimated \gls{mi} of the optimized input distribution and the incurred \gls{ber}. For comparison, we compare our results to known lower and upper bounds on capacity from the literature and show that the optimized \gls{mi} converges to the lower bound from the literature \cite{huleihel2021computable}. We also demonstrate the increase in \gls{mi} of the optimized input distribution in comparison to an \gls{iid} input distribution. To complete the experiments, we compare the \glspl{ber} incurred by the optimized input distribution and an \gls{iid} input distribution, showcasing an improvement of one order of magnitude for block lengths of $1024$.

\subsection{Contributions}

This paper presents the following contributions:

\begin{enumerate}
    \item Simultaneous input distribution optimization and code construction: We introduce a novel method that simultaneously maximizes input distributions and constructs practical codes through the application of \glspl{npd}.
    
    \item Neural polar decoders with SCL decoding: We demonstrate that \glspl{npd} can be effectively utilized with \gls{scl} decoding. This integration shows promising performance improvements, enhancing the practical utility of \glspl{npd}.
    
    \item Experimental validation and scalability: We validate our approach through extensive experiments, showing the quality of our method in optimizing \gls{mi} and reducing \glspl{ber} in comparison to uniform and \gls{iid} input distributions. Our results demonstrate its effectiveness for block lengths up to $1024$, suggesting that our method is scalable and potentially applicable to real-world communication systems.
\end{enumerate}

\subsection{Organization}
The remainder of this paper is organized as follows. Section \ref{sec:pre} provides the notations and necessary preliminaries. Section \ref{sec:npd} provides a comprehensive description of \glspl{npd}, including their structure and decoding process. It also introduces the key advancements made in this work, such as a fast estimation algorithm for \glspl{npd} parameters and their integration with \gls{scl} decoding. Section \ref{sec:input_opt} addresses the problem of optimizing input distributions to maximize the \gls{mi}. Section \ref{sec:applications} presents experimental results on capacity estimation and code design for various communication channels, including the Ising and Trapdoor channels. Finally, Section \ref{sec:conc} concludes the paper, summarizing the contributions and suggesting directions for future research.

\section{Notations and Preliminaries} \label{sec:pre}
\par Throughout this paper, we denote by $(\Omega, \cB, \bP)$ the underlying probability space upon which all random variables are defined. Here, $\Omega$ is the set of all two-sided infinite sequences of real numbers, represented by $\omega = (\ldots, \omega_{-1}, \omega_0, \omega_1, \ldots)$, with each $\omega_i \in \mathbb{R}$ for every integer $i$. The symbol $\cB$ denotes the corresponding Borel $\sigma$-algebra, ensuring that events involving these sequences are well-defined and measurable. The probability measure, denoted by $\bP$, is chosen to be the Lebesgue measure, which facilitates the assignment of probabilities to events within $\cB$. Lastly, $\bE$ denotes the expectation operator, used to calculate the expected value of random variables defined over this probability space.

\par\Glspl{rv} are denoted by capital letters and their realizations by lower-case letters, e.g., $X$ and $x$, respectively. Calligraphic letters denote sets, e.g., $\mathcal{X}$. We use the notation $X_i^j$ to denote the \gls{rv} $(X_i,X_{i+1},\dots,X_j)$, and $x_i^j$ to denote its realization for $i<j$. If $i=1$, we may omit the index $i$ to simplify notation, i.e., we use the shorthand $X^j$. The probability $\Pr[X = x]$ is denoted by $P_X(x)$. Stochastic processes are denoted by blackboard bold letters, e.g., $\bX := (X_i)_{i\in\bN}$. An $n$-coordinate projection of $\bP$ is denoted by $P_{X^n Y^n} := \bP\big|_{\sigma(X^n,Y^n)}$, where $\sigma(X^n,Y^n)$ is the $\sigma$-algebra generated by $(X^n,Y^n)$. For a given decomposition of $P_{X^n,Y^n}$ into $P_{X^n}$ and $P_{Y^n\vert X^n}$, we denote by $P_{X^n} \otimes P_{Y^n\vert X^n}$ the joint distribution obtained by $P_{X^n}(x^n) P_{Y^n\vert X^n}(y^n|x^n)$ for every $x^n, y^n \in \cX^n \times \cY^n$.

\par The \gls{mi} between two \glspl{rv} $X,Y$ is denoted by $\f{\sI}{X;Y}$. For two distributions $P,Q$, the \gls{ce} is denoted by $\ce{P}{Q}$, the entropy is denoted by $\f{\sH}{P}$ and the \gls{kl} divergence is denoted by $\kl{P}{Q}$. The notation $P\ll Q$ indicates that $P$ is absolutely continuous \gls{wrt} $Q$.
We denote by $[K:N]$ the set of integers $\{K,\dots,N\}$ for $K<N$.

\par The tuple $\left(W_{Y\vert X},\mathcal{X},\mathcal{Y}\right)$ defines a memoryless channel with input alphabet $\mathcal{X}$, output alphabet $\mathcal{Y}$ and a transition kernel $W_{Y\vert X}$. 
Throughout the paper we assume that $\mathcal{X} = \mathbb{F}_2$, where $\mathbb{F}_2$ is the Galois field with two elements, $\{0, 1\}$. 
For a memoryless channel, we denote its input distribution by $P_X = P_{X_i}$ for all $i\in\bZ$. {The tuple $\left(W_{Y\Vert X},\cX,\cY\right)$ defines a time-invariant channel with memory, where  $W_{Y\Vert X}=\left\{W_{Y_0\vert Y^{-1}_{-i+1},X^0_{-i+1}}\right\}_{i\in\bN}$, and $X_{-i}^0 = (X_{-i},X_{-i+1},\dots,X_{-1},X_0)$}. The term $W_{Y^N\Vert X^N}=\prod_{i=1}^{N}W_{Y_0\vert Y^{-1}_{-i+1},X^0_{-i+1}}$ denotes the probability of observing $Y^N$ causally conditioned on $X^N$ \cite{kramer1998directed}. The symmetric capacity of a channel is denoted by $\f{\sI}{W}$. {We denote by $\cD_{M,N}=\left\{x_{j,i},y_{j,i}\right\}_{j\in[M], i\in[N]} \sim P_{X^{MN}} \otimes W_{Y^{MN}\| X^{MN}}$ a finite sample of input-output pairs of $M$ consecutive blocks of $N$ symbols, where $x_{j,i}, y_{j,i}$ denotes the $i$-th input and output of the $j$-th block. The notation $x_{j,1}^N$ denotes  the collection $\{x_{j,i}\}_{i=1}^N$}. {The term $\cD_{MN}=\left\{x_{i},y_{i}\right\}_{i\in[MN]}$ denotes the same sample as $\cD_{M,N}$ after its concatenation into one long sequence of inputs and outputs pairs.}

\subsection{Polar Codes for Symmetric Channels}
{Let $G_N = B_N F^{\otimes n}$ represent the generator matrix for a block length of $N=2^n$, where $n \in \bN$, defining what is known as Ar\i{}kan's polar transform.}
The matrix $B_N$ is the permutation matrix associated with the bit-reversal permutation. It is defined by the recursive relation $B_N = R_N(I_2 \otimes B_{\frac{N}{2}})$ starting from $B_2=I_2$. The term $I_N$ denotes the identity matrix of size $N$ and $R_N$ denotes a permutation matrix called reverse-shuffle \cite{arikan2009channel}. {The matrix $G_N$ satisfies $G_NG_N=I_N$.}
The term $A\otimes B$ denotes the Kronecker product of $A$ and $B$. The term $A^{\otimes N}:=A\otimes A\otimes\dots\otimes A$ denotes an application of the $\otimes$ operator $N$ times. 

\par We define a polar code by the tuple $\left(\cX, \cY, W, E, F, G, H\right)$ that contains the channel $W$, the channel embedding $E$ and the core components of the \gls{sc} decoder, $F,G,H$. We define the {synthesized channels} by the tuple $\left(W^{(i)}_N,\cX,\cX^{i-1}\times\cY^N\right)$ for all $i\in[N]$. 
The term $E:\cY\to\cE$ denotes the channel embedding, where $\cE\subset\bR^d$. It is also referred in the literature by the term channel statistics, but here for our purposes, we alternatively choose to call it the channel embedding. For example, for a memoryless channel $W:=W_{Y\vert X}$, a valid choice of $E$, as used in the remainder of this paper, is given by the following:
\begin{equation}\label{eqn:memoryless-channel-stats}
    E(y) = \log\frac{\f{W}{y|1}}{\f{W}{y|0}} + \log\frac{\f{P_{X}}{1}}{\f{P_{X}}{0}},
\end{equation}
where the second term in the \gls{rhs} cancels out in the case where $P_X$ is uniform.
\par The functions $F:\cE\times\cE\to\cE, \;G:\cE\times\cE\times\cX\to\cE$ denote the check-node and bit-node operations, respectively. We denote by $H:\cE\to\bR$ a mapping of the embedding into an \gls{llr} value, i.e. a soft-decision. 
{For memoryless channels and with the selection of $E$ as defined in Equation \eqref{eqn:memoryless-channel-stats}, the functions $F$, $G$, and $H$ are specified as follows:
\begin{align}\label{eqn:sc_ops}
    &F(e_1, e_2) =  -2\tanh^{-1}\left(\tanh{\frac{e_1}{2}}\tanh{\frac{e_2}{2}}\right), \nn\\
    &G(e_1, e_2, u) =  e_2 + (-1)^{u}e_1, \nn\\
    &\hspace{0.1cm}H(e_1) = e_1,
\end{align}
where $e_1,e_2\in\bR, u\in\cX$}.
For this choice, the hard decision rule $h:[0,1]\to\cX$ is the sign function $h(l) = \mathbb{I}_{l>0.0}$, where $\bI$ is the indicator function. Applying \gls{sc} decoding on the channel outputs yields an estimate of the transmitted bits and their corresponding posterior distribution \cite{arikan2009channel}. 
{Specifically, after observing $y_1^N$, \gls{sc} decoding performs the map $$(y_1^N, f_1^N) \mapsto \left\{\hat{u}_i,\f{L_{U_i|U_1^{i-1},Y_1^N}}{\hat{u}_1^{i-1},y_1^N}\right\}_{i\in[N]},$$ where 
\begin{equation}\label{eqn:llr_def}
    \f{L_{U_i|U_1^{i-1},Y_1^N}}{u_1^{i-1},y_1^N} = \log\frac{\f{P_{U_i|U_1^{i-1},Y_1^N}}{1| u_1^{i-1},y_1^N}}{\f{P_{U_i|U_1^{i-1},Y_1^N}}{0| u_1^{i-1},y_1^N}}
\end{equation}
is the \gls{llr} function. The term $f_1^N$ represents the common information between the encoder and the decoder by setting 
\begin{equation*}
    f_i = \begin{cases} u_i & i\in\cA^c \\ 0.5 & i\in\cA \end{cases},
\end{equation*}
where the value $0.5$ is chosen arbitrarily to indicate that the bit needs to be decoded, the set $\cA\subseteq [N]$ is the information set and $\cA^c=[N]\setminus\cA$ is the frozen set. 
For more details on \gls{sc} decoding, the reader may refer to \cite[Section VIII]{arikan2009channel}.}

\subsection{SC List Decoding}\label{sec:list}
To enhance the error correction performance of polar codes, especially with codes of moderate lengths, the \gls{scl} decoding algorithm was introduced in \cite{tal2015list}. The fundamental concept behind list decoding lies in leveraging the structured nature of the polar transformation. Instead of relying solely on a single \gls{sc} decoder, the \gls{scl} decoder concurrently decodes multiple codeword candidates. This is achieved by applying multiple \gls{sc} decoders over the same channel's outputs, with the number of these decoders denoted as the list size $L$.

The \gls{scl} decoder generates a list of potential codewords, each ranked by its likelihood of being the transmitted message. 
To achieve this, the \gls{scl} algorithm estimates each bit's value ($0$ or $1$) while considering both possibilities. At each estimation step, the number of codeword candidates (also referred to as paths) doubles. To manage the algorithm's complexity, it employs a memory-saving strategy by retaining only a limited set of $L$ codeword candidates at any given time. 
To decide which paths to retain, a path metric is assigned to each path, determining its likelihood.
After the generation of $L$ paths, the \gls{scl} decoder chooses the codeword with the best path metric.

\subsection{Neural Networks}
\par {The class of shallow \glspl{nn}, i.e. \glspl{nn} with one hidden layer and with fixed input and output dimensions, is defined as follows \cite{schafer2006recurrent}.}
\begin{definition}[NN function class]\label{def:NN_function_class}
For the ReLU activation function
$\sigma_\mathsf{R}(x) = \max(x,0)$ and $d_i,d_o \in\bN$, define the class of neural networks with $k\in\bN$ neurons as:
\begin{align}
    \cG_\mathsf{NN}^{(d_i,k,d_o)}&:= \nn\\
    &\hspace{-1.5cm}\left\{g:\bR^{d_i}\to\bR^{d_o}: g(x)=\sum_{j=1}^k\beta_j \sigma_\sR( \mathrm{W}_j x+b_j),\ x\in\bR^{d_i} \right\},\label{eq:NN_def}
\end{align}
where $\sigma_\sR$ acts component-wise, $\beta_j\in\bR, \mathrm{W}_{j}\in \bR^{d_o \times d_i}$ and $ b_j\in\bR^{d_o}$ are the parameters of $g\in \cG_\mathsf{NN}^{(d_i,k,d_o)}$. 
Then, the class of NNs with input and output dimensions $(d_i,d_o)$ is given by
\begin{equation}
    \cG_\mathsf{NN}^{(d_i,d_o)} := \bigcup_{k\in\bN} \cG_\mathsf{NN}^{(d_i,k,d_o)},\label{eq:grnn_union}
\end{equation}
and the class of \glspl{nn} is given by $\cG_\mathsf{NN} := \bigcup_{d_i,d_o\in\bN} \cG_\mathsf{NN}^{(d_i,d_o)}$.
\end{definition}

\subsection{Communication Channels}\label{sec:applications:channels}

\par The channels selected in this study are chosen to validate our methodology through the experiments detailed in Section \ref{sec:applications}. Specifically, we focus on the \gls{awgn}, Ising, and Trapdoor channels. These channels serve as examples of both memoryless channels and \glspl{fsc}. The primary rationale behind selecting these channels is to assess the performance of the \gls{npd} against established theoretical benchmarks. For memoryless channels, we employ the \gls{sc} decoder \cite{arikan2009channel} as a theoretical benchmark, while for \glspl{fsc}, we use the \gls{sct} decoder as described in \cite{wang2015construction}. The definitions of the channels are provided below.

\par The \gls{awgn} channel is defined by the following model
$$ Y_i = X_i + N_i $$
where $N_i \stackrel{i.i.d.}{\sim} \cN(0,\sigma^2)$ and $X_i$ is the channel input subject to the average power constraint \(E[X_i^2] \leq P\).
The Ising channel \cite{ising1925beitrag} is an instance of \glspl{fsc} and is defined by
$$ Y_i = 
\begin{cases} 
X_i & \text{with probability } 0.5 \\
X_{i-1} & \text{with probability } 0.5. 
\end{cases}$$ 
The Trapdoor channel, introduced by Blackwell as a "two-state simple channel" \cite{blackwell1961}. It is defined by:
$$ Y_i = 
\begin{cases} 
X_i & \text{with probability } 0.5 \\
S_{i-1} & \text{with probability } 0.5, 
\end{cases}$$ 
and the channel state evolves according to the following deterministic rule $S_i = S_{i-1} + X_i + Y_i \mod 2$.

\section{Neural Polar Decoders}\label{sec:npd}
\par The purpose of this section is twofold. First, it aims to provide the necessary background on \glspl{npd}, as introduced in the authors' previous work \cite{aharoni2023data_arcive}. Second, it introduces two important advancements over the work presented in \cite{aharoni2023data_arcive}. These advancements include a fast estimation algorithm for the \gls{npd}'s parameters and its integration with list decoding \cite{tal2015list}. Specifically, Sections \ref{sec:npd:structure}, \ref{sec:npd:sc}, and \ref{sec:npd:consistency} contain material from the authors' previous work, whereas Sections \ref{sec:npd:scl} and \ref{sec:npd:est} present the new advancements.

\subsection{Structure of the NPD}\label{sec:npd:structure}
\par Let $X^N, Y^N$ be the channel inputs and outputs, respectively, and define $U^N=X^NG_N$. 
An \gls{npd} operates by decoding $Y^N$ into $\what{U}^N$ via a \gls{sc} procedure. 
The \gls{sc} procedure systematically decodes the information bits sequentially, utilizing the four core functions: embedding function $E$, check-node function $F$, bit-node function $G$, and embedding-to-LLR function $H$, which are repeatedly applied throughout the decoding process.
In an \gls{npd}, these four functions are approximated by \glspl{nn}. The subsequent definitions explain the \gls{npd} by describing its two primary blocks

\begin{definition}[Channel Embedding]\label{def:embedding}
    Let $E_{\theta_E}:\cY\to \bR^d$ be the \gls{nn} that embeds the channel outputs into the embedding space of the \gls{npd}, where $d\in\bN$ is the dimension of the embedding space. For $\cY \subset \bR$, we define $E_{\theta_E} \in \nnc{1}{k}{d}$ to be the embedding \gls{nn}, where $k\in\bN$ is the number of hidden units.
\end{definition}
\begin{definition}[Neural Polar Decoder]\label{def:npd}
Let $F_{\theta_F}$, $G_{\theta_F}$, $H_{\theta_H}$ be the \glspl{nn} components of the \gls{npd}. The \gls{npd}'s \glspl{nn} are defined by 
\begin{itemize}
    \item $F_{\theta_F} \in \nnc{2d}{k}{d}$ is the check-node \gls{nn},
    \item $G_{\theta_G} \in \nnc{2d+1}{k}{d}$ is the bit-node \gls{nn},
    \item $H_{\theta_H} \in \nnc{d}{k}{1}$ is the embedding-to-LLR \gls{nn},
\end{itemize}
where $d\in\bN$ is the dimension of the embedding space and $k\in\bN$ is the number of hidden units.
\end{definition}

\par In the rest of the paper, we adopt the notation $\theta = \{\theta_E,\theta_F,\theta_G,\theta_H\}$ for simplicity. The distinction between Definitions \ref{def:embedding} and \ref{def:npd} arises because the same \gls{npd} may be used for two different channel embeddings. This is illustrated in Section \ref{sec:input_opt}, where we use the same \gls{npd} to implement the \gls{hy} scheme \cite{honda2013polar}.

\subsection{SC Decoding with the NPD}\label{sec:npd:sc}
\par The \gls{npd}'s \gls{sc} decoder is defined by applying the recursive formulas of \gls{sc} decoding \cite{arikan2009channel} with $E_\theta$, $F_\theta$, $G_\theta$, and $H_\theta$ instead of $E,F,G,H$, as given in Equations \eqref{eqn:memoryless-channel-stats}, \eqref{eqn:sc_ops}. Specifically, upon observing the channel outputs, the function $E_\theta$ maps each channel output $Y_i$ into its embedding representation $e_{0,i}\in\bR^d$, acting as the belief of $X_i$ given $Y_i$. For example, $\mathbf{e}_0 = \left(e_{0,i}\right)_{i=1}^N \in\bR^{N\times d}$ represents the belief of $\mathbf{v}_0 := X^N$. Next, the \gls{npd} uses $\mathbf{e}_0$ as its input to decode $U^N$ sequentially by utilizing $F_\theta$, $G_\theta$, and $H_\theta$. Accordingly, for any $j \in [1:n]$ and $i \in [1:N]$, the recursive formulas are given by 
\begin{align}
    e_{j+1,2i-1} &= \f{F_\theta}{e_{j,i},\;e_{j,i+2^j}}, \\
    e_{j+1,2i} &= \f{G_\theta}{e_{j,i},\;e_{j,i+2^j},\;u_{j+1,\;2i-1}}.
\end{align}
The last function, $H_\theta$, converts an embedding $e_{j,i}$ into an \gls{llr} value associated with $v_{j,i}$, as defined by 
\begin{equation}
    l_{j,i} = \f{H_\theta}{e_{j,i}}, 
\end{equation}
where $l_{j,i}$ is the estimate of the \gls{llr} of $v_{j,i}$.
The \glspl{llr} $l_{j,i}$ are used to make decisions on the bits $v_{j,i}$ by applying the hard-decision function. The outputs of the \gls{sc} decoder are denoted by $\f{L^\theta_{U_i|U^{i-1}, Y^N}}{\hat{u}^{i-1},y^N}$ for $i\in[N]$.

\par The computational complexity of \gls{sc} decoding with the \gls{npd} is given next, as shown in \cite{aharoni2023data_arcive}.
\begin{theorem}[Computational Complexity of the \gls{npd}]\label{thm:nsc_complexity}
    Let $E_\theta\in\cG_\mathsf{NN}^{(1,k,d)},F_\theta\in\cG_\mathsf{NN}^{(2d,k,d)},G_\theta\in\cG_\mathsf{NN}^{(2d+1,k,d)}$ and $H_\theta\in\cG_\mathsf{NN}^{(d,k,1)}$. Then, the computational complexity of \gls{sc} decoding with the \gls{npd} is $\f{O}{kd N\log N}$.
\end{theorem}
\noindent The main purpose of Theorem \ref{thm:nsc_complexity} is to facilitate a comparison between the \gls{npd} and \gls{sct} decoder. Note that the computational complexity of the \gls{sct} decoder scales with the memory size of the channel as $O(|\cS|^3 N\log N)$ \cite{wang2015construction}. This highlights a key advantage of the \gls{npd}, as its computational complexity depends on the parameterization of the \gls{npd} and not on the channel's state size.

\subsection{SC List Decoding with the NPD}\label{sec:npd:scl}
\par We point out that list decoding \cite{tal2015list} can be easily integrated with the \gls{npd}. Recall that the \gls{npd} decoder uses the same structure as the \gls{sc} decoder and the \gls{sct} decoder, with the only distinction being the replacement of elementary operations with \glspl{nn}. Accordingly, we can seamlessly incorporate list decoding into the \gls{npd} decoder. Specifically, since the \gls{npd} decoding algorithm can estimate the \glspl{llr} at the decision points, we can leverage them to compute the path-metric and follow the same \gls{scl} decoding procedure. 
The following theorem shows the computational complexity of the \gls{npd} list decoder for the case where $E_\theta, F_\theta, G_\theta, H_\theta$ are \glspl{nn} with $k$ hidden units and the embedding space has $d$ dimensions. 
\begin{theorem}\label{thm:nscl_complexity}
    Let $E_\theta\in\cG_\mathsf{NN}^{(1,k,d)},F_\theta\in\cG_\mathsf{NN}^{(2d,k,d)},G_\theta\in\cG_\mathsf{NN}^{(2d+1,k,d)}$ and $ H_\theta\in\cG_\mathsf{NN}^{(d,k,1)}$. Then, the computational complexity of \gls{scl} decoding with the \gls{npd} and list size $L$ is $\f{O}{Lkd N\log N}$.
\end{theorem}

\subsection{Estimating the NPD's parameters}\label{sec:npd:est}
\par The parameters of the NPD are determined in a training phase. The goal of the training phase is to tune $\theta$ such that its outputs would match the true \glspl{llr} $\f{L_{U_i|U^{i-1},Y^N}}{y^N,u^{i-1}}, \;i\in[1:N]$, as defined in \eqref{eqn:llr_def}. For memoryless channels, the NPD is trained to approximate the ``vanilla'' \gls{sc} decoder \cite{arikan2009channel}; for \glspl{fsc}, the NPD is trained to approximate the \gls{sct} decoder \cite{wang2015construction}.
The training procedure of the NPD involves an iterative process. During each iteration, samples of channel input-output pairs are used to compute the gradient for updating $\theta$ via \gls{sgd} optimization. The training algorithm is given in Algorithm \ref{alg:npd_train}, and it is described hereafter.

\par Let $\mathcal{D}_{M,N} \sim P_{X^{MN}} \otimes W_{Y^{MN} \Vert X^{MN}}$ be a finite sample of input-output pairs consisting of $M$ consecutive blocks of $N$ symbols. At each iteration of the algorithm, a block $(x^N, y^N) \sim \mathcal{D}_{M,N}$ is drawn uniformly and used to compute $\f{\cL}{x^N,y^N;\theta}$, the loss for estimating the \gls{npd}'s parameters, as shown in \cite{aharoni2023data_arcive}. 
After computing the loss, the gradient $\nabla_\theta \f{\cL}{x^N,y^N;\theta}$ is computed to update the parameters of the \gls{npd} using \gls{sgd}.
This iterative scheme continues until the algorithm converges, either after a predetermined number of steps $\mathsf{N}_\mathsf{iter}$, or when $\f{\cL}{x^N,y^N;\theta}$ stops decreasing. The overall algorithm is outlined in Algorithm \ref{alg:npd_train}. 

\par In \cite{aharoni2023data_arcive}, the loss of the \gls{npd} is computed via the recursive formulas of the \gls{sc} decoding algorithm. Here, we enhance the gradient computation by calculating the loss stage-by-stage rather than through recursion. This improvement reduces the number of computational steps of the loss computation from $N \log(N)$ to $\log(N)$. Although the computational method has been modified, the guarantees of the algorithm remain intact because the underlying mathematical framework is preserved.
The details of the loss computation are described next.

\subsubsection{Loss Computation}
\par Given $x^N$ and $y^N$, the loss of the \gls{npd} is computed as follows. First, the embeddings of $\mathbf{v}_0 = x^N$ are computed as $\mathbf{e}_0 = \f{\mathbf{E}_\theta}{y^N}$, where $\mathbf{E}_\theta$ denotes the component-wise application of $E_\theta$ on each element of $y^N$. Similarly, the \glspl{llr} are computed as $\mathbf{l}_0 = \f{\mathbf{H}_\theta}{\mathbf{e}_0}$. The loss of the bits at stage $0$ is then computed by:
\begin{align}
    \f{\cL_0}{\mathbf{v}_0,\mathbf{e}_0;\theta} &= \nn\\
    &\hspace{-2cm}-\frac{1}{N}\sum_{i=1}^N v_{0,i}\log \f{\sigma}{l_{0,i}} + \overline{v}_{0,i} \log \overline{\f{\sigma}{l_{0,i}}},
\end{align}
where $\overline{x} = 1-x$, $\sigma(x) = \frac{1}{1+e^{-x}}$ is the logistic function and $\f{P_{U_i|U^{i-1},Y^N}}{1|u^{i-1},y^N}=\f{\sigma}{\f{L_{U_i|U^{i-1},Y^N}}{u^{i-1},y^N}}$.

\par At stage $1$, the loss is computed in the following manner. First, $\mathbf{v}_1$ is computed by 
\begin{equation}\label{eqn:gradient_v}
    \mathbf{v}_1 = [\mathbf{v}_0^o \oplus \mathbf{v}_0^e \mid \mathbf{v}_0^e],
\end{equation}
where $\mathbf{v}_0^o,\mathbf{v}_0^e\in \bF_2^{\frac{N}{2}}$ contain the odd and even elements of $\mathbf{v}_0$, respectively. Next, $\mathbf{e}_1$ is computed by:
\begin{align}\label{eqn:gradient_e}
    \mathbf{e}_1 = [\f{\mathbf{F}_\theta}{\mathbf{e}_0^o,\mathbf{e}_0^e} \mid 
                \f{\mathbf{G}_\theta}{\mathbf{e}_0^o,\mathbf{e}_0^e, \mathbf{v}_0^o \oplus \mathbf{v}_0^e} ],
\end{align}
where $\f{\mathbf{F}_\theta}{\mathbf{e}_0^o,\mathbf{e}_0^e},\f{\mathbf{G}_\theta}{\mathbf{e}_0^o,\mathbf{e}_0^e, \mathbf{v}_0^o \oplus \mathbf{v}_0^e}\in\bR^{\frac{N}{2}\times d}$, the operator $[\cdot|\cdot]$ denotes the concatenation of two matrices along the first dimension, and $\mathbf{e}_1\in\bR^{N\times d}$. The loss of stage $1$ is then computed by first computing:
\begin{equation}\label{eqn:gradient_l}
\mathbf{l}_1 = \f{\mathbf{H}_\theta}{\mathbf{e}_1}, 
\end{equation}
and then computing $\f{\cL_1}{\mathbf{v}_1,\mathbf{e}_1;\theta}$, as done in stage $0$.

\par At stages $j\in[2:n]$, the same computations as in stage $1$ are followed, but they are performed within sub-blocks independently. Given $\mathbf{e}_{j-1}$ and $\mathbf{v}_{j-1}$, we first split them into the collections:
\begin{align}\label{eqn:gradient_split}
    \mathbf{v}^{(j-1)}&=\left\{\mathbf{v}_{j-1,k}\right\}_{k=0}^{2^{j-1}-1} \nn\\
    \mathbf{e}^{(j-1)}&=\left\{\mathbf{e}_{j-1,k}\right\}_{k=0}^{2^{j-1}-1},
\end{align}
where 
\begin{align}\label{eqn:gradient_split}
    \mathbf{v}_{j-1,k} &= \left\{v_{j-1,l}\right\}_{l=1+k2^{n-j+1}}^{(k+1)2^{n-j+1}} \nn\\
    \mathbf{e}_{j-1,k} &= \left\{e_{j-1,l}\right\}_{l=1+k2^{n-j+1}}^{(k+1)2^{n-j+1}}, \nn
\end{align}
with $\mathbf{v}_{j-1,k} \in \bF_2^{\frac{N}{2^{j-1}}\times 1}$ and $\mathbf{e}_{j-1,k} \in \bR^{\frac{N}{2^{j-1}}\times d}$.
For every $\mathbf{v}_{j-1,k}$ and $\mathbf{e}_{j-1,k}$, we repeat the computations in Equations \eqref{eqn:gradient_v}--\eqref{eqn:gradient_e} to produce $\mathbf{v}^\prime_{j-1,k}$ and $\mathbf{e}^\prime_{j-1,k}$, which are then concatenated into $\mathbf{e}_{j}$, $\mathbf{v}_{j}$. Next, $\mathbf{e}_{j}$, $\mathbf{v}_{j}$ are used to compute $\cL_j\left(\mathbf{v}_j,\mathbf{e}_j;\theta\right)$ and are passed to the next stage. The overall loss is computed by 
\begin{equation}\label{eqn:npd_loss}
    \f{\cL}{x^n,y^n;\theta} = \frac{1}{n+1}\sum_{j=0}^n \f{\cL_j}{\mathbf{v}_j,\mathbf{e}_j;\theta},
\end{equation}
and the corresponding gradient is given by $\nabla_\theta \f{\cL}{x^n,y^n;\theta}$. The loss computation is summarized in Algorithm \ref{alg:npd_loss}. 


\begin{algorithm}[!t]
\caption{Loss Computation for Neural Polar Decoders}
\begin{algorithmic}

\State \textbf{Input:} 
\Statex \hspace{1em} $x^N,y^N$ \Comment{Channel input-output pairs}
\Statex \hspace{1em} $E_\theta, F_\theta, G_\theta, H_\theta$ \Comment{\gls{npd} model}
\State \textbf{Output:} 
\Statex \hspace{1em} $\mathcal{L}(x^N, y^N; \theta)$ \Comment{computed loss}
\algrule
\State \textbf{Stage 0:}
    \State \hspace{1em}$\mathbf{v}_0 \leftarrow x^N$
    \State \hspace{1em}$\mathbf{e}_0 \leftarrow \f{\mathbf{E}_\theta}{y^N}$
    \State \hspace{1em}$\mathbf{l}_0 \leftarrow \f{\mathbf{H}_\theta}{\mathbf{e}_0}$
    \vspace{-2em}
    \State \begin{align*}
    \hspace{0.3em}\mathcal{L}_0(\mathbf{v}_0, \mathbf{e}_0; \theta)= -\frac{1}{N} \sum_{i=1}^N v_{0,i} \log \sigma(l_{0,i}) + \overline{v}_{0,i} \log \overline{\sigma(l_{0,i})} 
    \end{align*}
        \Comment{loss of stage $0$}

\State \textbf{Stage 1:}
    \State \hspace{1em}$\mathbf{v}_1 \gets [\mathbf{v}_0^o \oplus \mathbf{v}_0^e \mid \mathbf{v}_0^e]$
    \State \hspace{1em}$\mathbf{e}_1 \gets [\mathbf{F}_\theta(\mathbf{e}_0^o, \mathbf{e}_0^e) \mid \mathbf{G}_\theta(\mathbf{e}_0^o, \mathbf{e}_0^e, \mathbf{v}_0^o \oplus \mathbf{v}_0^e)]$
    \State \hspace{1em}$\mathbf{l}_1 \gets \mathbf{H}_\theta(\mathbf{e}_1)$
        \vspace{-2em}
    \State \begin{align*}
    \hspace{0.3em}\mathcal{L}_1(\mathbf{v}_1, \mathbf{e}_1; \theta)= -\frac{1}{N} \sum_{i=1}^N v_{1,i} \log \sigma(l_{1,i}) + \overline{v}_{1,i} \log \overline{\sigma(l_{1,i})} 
    \end{align*}
    \Comment{loss of stage $1$}
    
\State \textbf{Stages 2 to $\mathbf{n}$:}
\hspace{1em}\For{$j = 2$ to $n$}
    \State Compute $\mathbf{v}^{(j-1)}$ and $\mathbf{e}^{(j-1)}$ \Comment{Equation \eqref{eqn:gradient_split}}
    \State Initiate $\mathbf{v}^{(j)}=\emptyset, \mathbf{e}^{(j)}=\emptyset, \mathbf{l}^{(j)}=\emptyset$
    \For{each $\mathbf{\tilde{v}}_{j-1} \in \mathbf{v}^{(j-1)}$ and $\mathbf{\tilde{e}}_{j-1} \in \mathbf{e}^{(j-1)}$}
        \State Compute $\mathbf{\tilde{v}}_j$, $\mathbf{\tilde{e}}_j$ and $\mathbf{\tilde{l}}_j$ \Comment{Equation \eqref{eqn:gradient_v}--\eqref{eqn:gradient_l}} 
        \State $\mathbf{v}^{(j)}\gets\mathbf{v}^{(j)}\cup\mathbf{\tilde{v}}_j$, $\mathbf{e}^{(j)}\gets\mathbf{e}^{(j)}\cup\mathbf{\tilde{e}}_j$, $\mathbf{l}^{(j)}\gets\mathbf{l}^{(j)}\cup\mathbf{\tilde{l}}_j$
    \EndFor
    \State Concatenate $\mathbf{v}^{(j)},\mathbf{e}^{(j)},\mathbf{l}^{(j)}$ into $\mathbf{e}_j$, $\mathbf{v}_j$, $\mathbf{l}_j$ 
        \vspace{-2em}
    \State \begin{align*}
    \hspace{0.3em}\mathcal{L}_j(\mathbf{v}_j, \mathbf{e}_j; \theta)= -\frac{1}{N} \sum_{i=1}^N v_{j,i} \log \sigma(l_{j,i}) + \overline{v}_{j,i} \log \overline{\sigma(l_{j,i})} 
    \end{align*}
    \Comment{loss of stage $j$}
    
\EndFor

\State \Return $\mathcal{L}(x^N, y^N; \theta) \gets \frac{1}{n+1} \sum_{j=0}^n \mathcal{L}_j(\mathbf{v}_j, \mathbf{e}_j; \theta)$

\end{algorithmic}\label{alg:npd_loss}
\end{algorithm}

\subsection{Consistency of the NPD Estimation}\label{sec:npd:consistency}
\par In \cite{aharoni2023data_arcive}, Algorithm \ref{alg:npd_train} was shown to be consistent. That is, as $M$ approaches infinity, the optimized \gls{npd} recovers the conditional entropies of the effective channels. The consistency of the \gls{npd} \cite[Theorem 4]{aharoni2023data_arcive} is given herein.
\begin{theorem}\label{thm:npd_consistent} 
    Let $\bX,\bY$ be the inputs and outputs of an indecomposable \gls{fsc}. 
    Let $\cD_{M,N} \sim P_{X^{MN}} \otimes W_{Y^{MN}\| X^{MN}}$, where $N=2^n,\;M,n\in \bN$. 
    Let  $u_{j,i} = (x_{j,1}^{N} G_N)_i$. Then, for every $\varepsilon>0$ there exists $p\in\bN$, compact $\Theta \in\bR^p$ and $m\in\bN$ such that for $M>m$ and $i\in[N]$, $\bP-a.s.$
    \begin{align}
        \left| \f{\sH_{\Theta}^M}{U_i\vert U^{i-1}, Y^N} -\f{\sH}{U_i\vert U^{i-1}, Y^N}\right| < \varepsilon,
    \end{align}
    where 
    \begin{align}\label{eqn:ce_nsc_loss}
        \f{\sH_{\Theta}^M}{U_i\vert U^{i-1}, Y^N} &= \min_{\theta\in\Theta} \left\{ \frac{1}{M}\sum_{j=1}^M -\log\f{\sigma}{\f{l_\theta}{y^N,u^i}}\right\}.
    \end{align}
\end{theorem}
\begin{algorithm}[t!]
    \caption{Data-driven \gls{npd} Estimation}
    \label{alg:npd_train}
    \textbf{input:} 
    Dataset $\cD_{M,N}$, \#of iterations $\mathsf{N_{iters}}$, learning rate $\gamma$ \\
    \textbf{output:} Optimized $\theta^\ast$
    \algrule
    \begin{algorithmic}
    \State Initiate the weights of $E_\theta, F_\theta, G_\theta, H_\theta$
    \For{$\mathsf{N_{iters}}$ iterations} 
        \State Sample $x^N, y^N\sim \cD^\psi_{M,N}$
        \State Compute $\f{\cL}{x^N,y^N;\theta}$ \Comment{Equation \eqref{eqn:npd_loss}}
        \State Update $\theta := \theta - \gamma \nabla_\theta \f{\cL}{x^N,y^N;\theta}$ 
    \EndFor \\
    \Return $\theta^\ast$ 
    \end{algorithmic}
\end{algorithm}

\section{Input Distribution Optimization}\label{sec:input_opt}
\par This section addresses the problem of choosing an input distribution $P^\psi_{X^N}$ that maximizes $\f{\sI}{X_\psi^N;Y^N}$. From this section onwards, the subscript $\psi$ designates the dependence of the \gls{mi} on the specific input distribution parameterized by $\psi$. The process of determining an input distribution that maximizes $\f{\sI}{X_\psi^N;Y^N}$ is composed of two main steps. In the first step, the input distribution is fixed. For a fixed $\psi$, Algorithm \ref{alg:npd_train} is employed to estimate $\f{\sI}{X_\psi^N;Y^N}$. In the second step, the parameters of the \gls{npd} are fixed. For a fixed \gls{npd}, the gradient of the input distribution $\psi$ is computed. Together, these two steps complement each other and are applied interchangeably to form an alternated maximization procedure. This completes the overview of the rate optimization scheme that is detailed herein.
Section \ref{sec:input_opt:model} describes the parametric model we used for $P^\psi_{X^N}$. Section \ref{sec:mi_est} illustrates the estimation step. Section \ref{sec:mi_max} describes the optimization step. Section \ref{sec:mi_alg} concludes with the overall algorithm.

\subsection{Parametric model of the input distribution} \label{sec:input_opt:model}
\par The input distribution is parameterized using a \gls{lstm} model \cite{hochreiter1997long}, as detailed herein.
Let $\psi = \{\psi_1,\psi_2\}$ be the parameters of the following model:
\begin{align}\label{eqn:input_model}
    [h_{i+1}, c_{i+1}] &= \mathsf{LSTM}_{\psi_1}(x_{i}, h_{i}, c_{i}) \nn\\
    o_{i}  &= W_{\psi_2} h_{i}+ b_{\psi_2},
\end{align}
where $\mathsf{LSTM}_{\psi_1}$ acts as a kernel that takes as an input $x_{i} \in \bF_2$ and current states $ h_{i}, c_{i}$, and outputs its next inner state and outer state, $c_{i+1}, h_{i+1} \in \bR^k$, respectively. The initial states $c_1, h_1 \in \bR^k$ are set arbitrarily as zero vectors. A linear transformation is then applied to $h_{i}$ using $W_{\psi_2} \in \bR^{1 \times k}$ and $ b_{\psi_2} \in \bR$. In Equation \eqref{eqn:input_model}, the output $o_{i}$ forms the \gls{llr} at time $i$ denoted by $L^\psi_{X_i|X^{i-1}}(x^{i-1})$, as it is a function of $x^{i-1}$ through the states $c_{i}$ and $h_{i}$.

\par Given $\psi$, sampling a sequence $x^N \sim P^\psi_{X^N}$ involves the following steps. Let $A^N$ be a sequence of \gls{iid} \glspl{rv} with $A_1\sim \mathsf{Unif}[0,1]$.
For every $i \ge 1$, $x_i$ is sampled according to the following rule:
\begin{align}\label{eqn:sampling_xi}
o_{i} &= W_{\psi_2} h_{i} + b_{\psi_2} \nn\\
x_i &= \mathbb{I}(A_{i} < \sigma(o_{1})) \nn\\
[h_{i+1}, c_{i+1}] &= \mathsf{LSTM}_{\psi_1}(x_{i}, h_{i}, c_{i}).
\end{align}
where $\sigma(x) = \frac{1}{1+e^{-x}}$ is the logistic function that maps $\f{L^\psi_{X_i|X^{i-1}}}{x^{i-1}}$ into $P^\psi_{X_i|X^{i-1}}(1|x^{i-1})$.
Thus, by using the model in Equation \eqref{eqn:input_model} and sampling $x^N$ by Equation \eqref{eqn:sampling_xi}, we obtain $x^N\sim P^{\psi}_{X^N}$. 
The log-likelihood is then evaluated as follows:
\begin{equation}\label{eqn:input_model_joint}
    \log \f{P^\psi_{X^N}}{x^N} = \sum_{i=1}^N\Big(x_i\log\f{\sigma}{o_i}+ \bar{x}_i\log\overline{\f{\sigma}{o_i}}\Big).
\end{equation}

\subsection{Step 1: MI Estimation}\label{sec:mi_est}
The estimation step considers a fixed input distribution $P_{X^N}^\psi$, and a channel $W_{Y^N\|X^N}$. These distributions define the joint distribution $P^\psi_{X^N,Y^N} = P_{X^N}^\psi \otimes W_{Y^N\|X^N}$ and the corresponding \gls{mi} $\f{\sI}{X_\psi^N;Y^N}$. Since $U_\psi^N=X_\psi^NG_N$ is bijective, it follows that $\f{\sI}{X_\psi^N;Y^N}=\f{\sI}{U_\psi^N;Y^N}$. By the factorization of the \gls{mi} as a difference of conditional entropies, we have:
\begin{equation}\label{eqn:mi_doe}
    \f{\sI}{U_\psi^N;Y^N} = \f{\sH}{U_\psi^N}-\f{\sH}{U_\psi^N|Y^N}. 
\end{equation}

\par Equation \eqref{eqn:mi_doe} suggests that the \gls{mi} $\f{\sI}{U_\psi^N;Y^N}$ can be estimated by employing the \gls{npd} for each term in the \gls{rhs} of Equation \eqref{eqn:mi_doe}. 
Let an \gls{npd} be defined by the \glspl{nn} $F_{\theta_F}, G_{\theta_G}, H_{\theta_H}$, as given in Definition \ref{def:npd}. Let $E_{\theta_{E_1}}, E^{\mathsf{co}}_{\theta_{E_2}}$ be two channel embeddings, as given in Definition \ref{def:embedding}. The \gls{nn} $E_{\theta_{E_1}}$ embeds the channel outputs to yield $\mathbf{e}_0 = \f{\mathbf{E}_{\theta_{E_1}}}{y^N}$, while the \gls{nn} $E_{\theta_{E_2}}$ produce the embeddings for $\f{\sH}{U^N}$ by embedding constant inputs to yield $\mathbf{e}^\mathsf{co}_0 = \f{\mathbf{E}_{\theta_{E_2}}}{\mathbf{0}^N}\in \bR^{N\times d}$. To simplify notation, we define the parameters associated with applying the \gls{npd} \textit{after} observing the channel outputs by $\theta^\mathsf{ch}=\{E_{\theta_{E_1}},F_{\theta_F}, G_{\theta_G}, H_{\theta_H}\}$; the parameter associated with applying the \gls{npd} \textit{before} observing the channel outputs by $\theta^\mathsf{co}=\{E_{\theta_{E_2}},F_{\theta_F}, G_{\theta_G}, H_{\theta_H}\}$, and all the parameters by $\theta = \{\theta^\mathsf{ch},\theta^\mathsf{co}\}$. 

\par Application of Algorithm \ref{alg:npd_train} with $\cD^{\psi}_{M,N}\sim P^\psi_{X^{MN}} \otimes W_{Y^{MN}\| X^{MN}}$ as inputs returns $\theta^\mathsf{ch}$ that enables the estimation of $\f{\sH}{U_\psi^N|Y^N}$. The estimate is given by
\begin{equation}
\f{\sH_{\theta^\mathsf{ch}}^M}{U^N_\psi\vert Y^N} =  \frac{N}{M}\sum_{x^N,y^N \sim \cD^{\psi}_{M,N}} \f{\cL}{x^N,y^N;\theta^\mathsf{ch}}.
\end{equation}
In the same manner, application of Algorithm \ref{alg:npd_train} with $\cD^{\psi}_{M,N}\sim P^\psi_{X^{MN}} \otimes \bI_{Y=0}^{\otimes MN}$ as inputs returns $\theta^\mathsf{co}$ that enables the estimation of $\f{\sH}{U_\psi^N}$. Here, $\mathbb{I}_{Y = 0}$ denotes a constant distribution that assigns the value $0$ with probability $1$. Accordingly, the estimate of $\f{\sH}{U_\psi^N}$ is given by
\begin{equation}
\f{\sH_{\theta^\mathsf{co}}^M}{U^N_\psi} =  \frac{N}{M}\sum_{x^N,\mathbf{0}^N \sim \cD^{\psi}_{M,N}} \f{\cL}{x^N,\mathbf{0}^N;\theta^\mathsf{co}}.
\end{equation}
Together, the parameters $\theta^\mathsf{ch}$ and $\theta^\mathsf{co}$ form the parameters for the estimation $\f{\sI_\psi}{U^N;Y^N}$ as given by
\begin{align}\label{eqn:mi_est_step}
    \f{\what{\sI}^M_{\Theta}}{U_\psi^N;Y^N} &=\min_{\theta^\mathsf{co}\in\Theta}
    \frac{N}{M}\sum_{x^N,y^N \sim \cD^{\psi}_{M,N}}  \f{\cL}{x^N,\mathbf{0}^N;\theta^\mathsf{co}} -  \nn\\
    &\min_{\theta^\mathsf{ch}\in\Theta}
    \frac{N}{M}\sum_{x^N,y^N \sim \cD^{\psi}_{M,N}}\f{\cL}{x^N,y^N;\theta^\mathsf{ch}}.
\end{align}
We also denote by $\theta^\ast = \{\theta^{\ast\mathsf{co}},\theta^{\ast\mathsf{ch}}\}$ the argument achieving the optimum of Equation \eqref{eqn:mi_est_step}.
The following theorem identifies that $\f{\what{\sI}^M_{\Theta}}{U_\psi^N;Y^N}$ is a consistent estimator of $\f{\sI}{U_\psi^N;Y^N}$.
\begin{theorem}\label{thm:npd_mi_consistent}
    Let $\bX,\bY$ be the inputs and outputs of an indecomposable \gls{fsc}. 
    Let $\cD^\psi_{M,N} \sim P^\psi_{X^{MN}} \otimes W_{Y^{MN}\| X^{MN}}$, where $N=2^n,\;M,n\in \bN$. 
    Then, for every $\varepsilon>0$ there exists $p\in\bN$, compact $\Theta \in\bR^p$ and $m\in\bN$ such that for $M>m$ and $i\in[N]$, $\bP-a.s.$
    \begin{align}
        \left| \f{\what{\sI}^M_{\Theta}}{U_{\psi,i};Y^N|U^{i-1}_\psi} -\f{\sI}{U_{\psi,i};Y^N|U^{i-1}_\psi}\right| < \varepsilon.
    \end{align}
\end{theorem}
\noindent The proof of the theorem follows the consistency of the \gls{npd} in Theorem \ref{thm:npd_consistent} and the triangle inequality. 
\begin{proof}
    For the proof we use Theorem \ref{thm:npd_consistent} to claim that for every $i\in[1:N]$ we have the consistency of $\f{\sH_{\Theta}^M}{U_i\vert U^{i-1}}$ and $\f{\sH_{\Theta}^M}{U_i\vert U^{i-1}, Y^N}$. Hence, we have that for any $\frac{\varepsilon}{2}$, $\bP-a.s.$
    \begin{align*}
        &\left| \f{\sH_{\Theta}^M}{U_i\vert U^{i-1}, Y^N} -\f{\sH}{U_i\vert U^{i-1}, Y^N}\right| < \frac{\varepsilon}{2}, \\
        &\left| \f{\sH_{\Theta}^M}{U_i\vert U^{i-1}} -\f{\sH}{U_i\vert U^{i-1}}\right| < \frac{\varepsilon}{2}. 
    \end{align*}
    Thus, due to the triangle inequality, we obtain that $\bP-a.s.$
    \begin{align*}
        \left| \f{\what{\sI}^M_{\Theta}}{U_{\psi,i};Y^N|U^{i-1}_\psi} -\f{\sI}{U_{\psi,i};Y^N|U^{i-1}_\psi}\right| &\leq  \\
        &\hspace{-6cm}\left| \f{\sH_{\Theta}^M}{U_i\vert U^{i-1}} -\f{\sH}{U_i\vert U^{i-1}}\right| + \\
        &\hspace{-6cm}\left| \f{\sH_{\Theta}^M}{U_i\vert U^{i-1}, Y^N} -\f{\sH}{U_i\vert U^{i-1}, Y^N}\right| < \frac{\varepsilon}{2}+\frac{\varepsilon}{2}=\varepsilon.
    \end{align*}
\end{proof}

\subsection{Step 2: MI maximization}\label{sec:mi_max}
\par The maximization step involves fixing the parameter $\theta^\ast$ obtained from the estimation step and maximizing $\f{\what{\sI}^M_{\Theta}}{U_\psi^N;Y^N}$ \gls{wrt} $\psi$. The following theorem describes the gradient of $\f{\sI}{U_\psi^N;Y^N}$ \gls{wrt} $\psi$.
\begin{theorem}\label{thm:grads_without_oracle}
    The gradient of $\f{\sI}{U_\psi^N;Y^N}$ \gls{wrt} $\psi$ is given by 
    \begin{align}\label{eqn:mi_grad}
        \nabla_\psi \f{\sI}{U_\psi^N;Y^N} &= \nn\\
        &\hspace{-1.5cm}\bE_{P_{X^N}^\psi\otimes W_{Y^N\Vert X^N}}\Bigg[\nabla_\psi\log \f{P_{X^N}^\psi}{X^N}
        \f{Q}{X^N,Y^N}\Bigg],
    \end{align}
    where $\f{Q}{X^N,Y^N} = \log\frac{\f{P_{U^N|Y^N}}{X^NG_N\vert Y^N}}{\f{P_{U^N}}{X^NG_N}}$.
\end{theorem}
\par Theorem \ref{thm:grads_without_oracle} states that the gradients of the input distribution parameters $\psi$ are computed via a re-parameterization trick \cite{schulman2015gradient}, which allows for the computation of gradients through samples drawn from the joint distribution. The proof of Theorem \ref{thm:grads_without_oracle} is given next.
\begin{algorithm}[b!]
    \caption{Code Rate Optimization via Neural Polar Decoders}
    \label{alg:npd_mi_max}
    \textbf{input:} 
    Channel $W$, block length $N$, \#iterations $\mathsf{N_{iters}}$, \#iterations of warm-up $\mathsf{N_{warm-up}}$, learning rate $\gamma$ \\
    \textbf{output:} Optimized $\theta^\ast,\psi^\ast$
    \algrule
    \begin{algorithmic}
    \State Initiate the weights of $\theta,\psi$
    \State Generate $\cD^\psi_{M,N} \sim P^\psi_{X^N} \otimes W_{Y^N\|X^N}$
    \State \underline{\textbf{Warm-up phase:}}
    \For{$\mathsf{N_{warm-up}}$ iterations} 
        \State Sample $x^N, y^N\sim \cD^\psi_{M,N}$
        \State Update 
        \vspace{-0.5em}\begin{align*}
            \quad\theta := \theta - \gamma \nabla_\theta \Big[\f{\cL}{x^N,y^N;\theta^\mathsf{ch}} + \f{\cL}{x^N,\mathbf{0}^N;\theta^\mathsf{co}}\Big] 
        \end{align*}
    \EndFor \\
    \State \underline{\textbf{Main phase:}}
    \For{$\mathsf{N_{iters}}$ iterations} 
        \State Sample $x^N, y^N\sim \cD^\psi_{M,N}$
        \State \textbf{Maximization step:}
        \State \hspace{1em}Update $\psi := \psi + \gamma\nabla_\psi \f{\what{\sI}_{\theta}}{U_\psi^N;Y^N} $ 
        \State \hspace{1em}Generate $\cD^\psi_{M,N} \sim P^\psi_{X^N} \otimes W_{Y^N\|X^N}$
        \State \textbf{Estimation step:} 
        \State \hspace{1em}Sample $x^N, y^N\sim \cD^\psi_{M,N}$
        \State \hspace{1em}Update 
        \vspace{-0.5em}\begin{align*}
            \quad\quad\quad\theta := \theta - \gamma \nabla_\theta \Big[\f{\cL}{x^N,y^N;\theta^\mathsf{ch}} + \f{\cL}{x^N,\mathbf{0}^N;\theta^\mathsf{co}}\Big] 
        \end{align*}
    \EndFor \\
    \Return $\theta, \psi$ 
    \end{algorithmic}
\end{algorithm}
\begin{proof}\label{prf:grads_without_oracle}
    Let $\f{Q}{X^N,Y^N} = \log\frac{\f{P_{U^N|Y^N}}{X^NG_N\vert Y^N}}{\f{P_{U^N}}{X^NG_N}}$. The gradient of $\f{\sI}{U_\psi^N;Y^N}$ \gls{wrt} $\psi$ is computed by the following steps: 
    \begin{align*}
        \nabla_\psi \f{\sI}{U_\psi^N;Y^N} \\
        &\hspace{-2cm}= \nabla_\psi \hspace{-0.8cm}\sum_{x^N,y^N \in \cX^N\times\cY^N}\hspace{-0.75cm}\f{P_{X^N}^\psi}{x^N}\f{W_{Y^N\Vert X^N}}{y^N\| x^N}\f{Q}{x^N,y^N}  \\
        &\hspace{-2cm}= \hspace{-0.8cm}\sum_{x^N,y^N\in \cX^N\times\cY^N}\hspace{-0.75cm}\nabla_\psi \f{P_{X^N}^\psi}{x^N}\f{W_{Y^N\Vert X^N}}{y^N\| x^N}\f{Q}{x^N,y^N}  \\
        &\hspace{-2cm}=\Ep{P_{X^N}^\psi\otimes W_{Y^N\Vert X^N}}{\nabla_\psi\log \f{P_{X^N}^\psi}{X^N} \f{Q}{X^N,Y^N} }.
    \end{align*}
    In the last equality, we multiplied and divided by $\f{P_{X^N}^\psi}{x^N}$, and used the fact the $\frac{\dd}{\dd x} \log f(x) = \frac{\frac{\dd}{\dd x}f(x)}{f(x)}$.
\end{proof}
\par Practically, the algorithm exploits the consistency of the \glspl{npd} to substitute $\log\frac{\f{P_{U^N|Y^N}}{X^NG_N\vert Y^N}}{\f{P_{U^N}}{X^NG_N}}$ with $\f{\cL}{X^N,\mathbf{0}^N;\theta^{\ast\mathsf{co}}} - \f{\cL}{X^N,Y^N;\theta^{\ast\mathsf{ch}}}$ as plug-in estimator. Thus, given $\theta^\ast$, the gradients used to optimize $\psi$ are given by:
\begin{align}\label{eqn:mi_grad_est}
    \nabla_\psi \f{\what{\sI}_{\theta^\ast}}{U_\psi^N;Y^N} &= \bE_{P_{X^N}^\psi\otimes W_{Y^N\Vert X^N}}\Bigg[ \nabla_\psi\log \f{P_{X^N}^\psi}{X^N} \times \nn\\ 
    &\hspace{-1.5cm}\Big[\f{\cL}{X^N,\mathbf{0}^N;\theta^{\ast\mathsf{co}}} - \f{\cL}{X^N,Y^N;\theta^{\ast\mathsf{ch}}}\Big] \Bigg].
\end{align}

\subsection{Overall Algorithm}\label{sec:mi_alg}
\par The overall algorithm integrates the two steps outlined in Sections \ref{sec:mi_est}, \ref{sec:mi_max} through an alternating maximization procedure. The algorithm begins with a ``warm-up" phase, during which only step 1 is repeated. This initial phase focuses on estimating $\f{\sI}{U_\psi^N;Y^N}$, as its estimate is subsequently utilized as a proxy for $\log\frac{P_{U^N|Y^N}}{P_{U^N}}$ in \eqref{eqn:mi_grad}. Following the warm-up phase, step 1 and 2 are repeated interchangeably. The complete algorithm is detailed in Algorithm \ref{alg:npd_mi_max}.

\section{Capacity Estimation and Code Design}\label{sec:applications}

\begin{figure}[b!]
    \centering
        \input{figures/awgn_mi.tex}         
        \caption{Evolution of $\f{\sI_\psi}{U^N;Y^N}$ with respect to the iteration number of Algorithm \ref{alg:npd_mi_max} for the binary-input \gls{awgn} channel. }
        \label{fig:awgn_mi}
\end{figure}
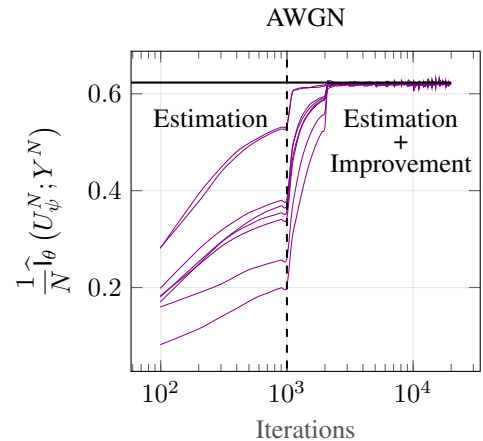
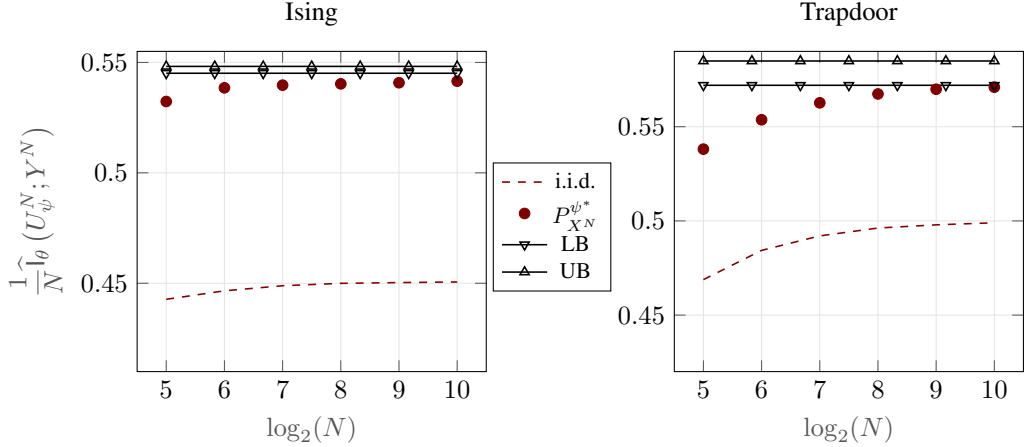
\begin{figure*}[!t]
    \centering
    \input{figures/nsc-optimize}
    \caption{Estimated \gls{mi} per symbol for the Ising and Trapdoor channels with various block lengths.}
    \label{fig:fscs_mi}
\end{figure*}
\par This section highlights the utility of \glspl{npd} in estimating channel capacity and designing codes with optimized code rates by applying Algorithm \ref{alg:npd_mi_max}. It demonstrates the dual capabilities of the outcome of Algorithm \ref{alg:npd_mi_max} in addressing both tasks simultaneously. We present results for the Ising and Trapdoor channels, leveraging existing capacity bounds from the literature. Additionally, we use the \gls{sct} decoder with \gls{iid} uniform inputs to provide a theoretical benchmark for comparing with our algorithm. Sections \ref{sec:applications:capacity} and \ref{sec:applications:coding} investigate the outcomes of Algorithm \ref{alg:npd_mi_max}, focusing on the optimized input distribution with parameters $\psi^\ast$ and the \gls{npd} with parameters $\theta^\ast$. Section \ref{sec:applications:capacity} evaluates the estimated value of the optimized \gls{mi}. Section \ref{sec:applications:coding} examines the \glspl{ber} achieved by the optimized input distribution and \gls{npd}.
\begin{figure*}[b!]
    \begin{minipage}{0.49\textwidth}
        \centering
        \includegraphics[width=\columnwidth, trim={0 0 0 55},clip]{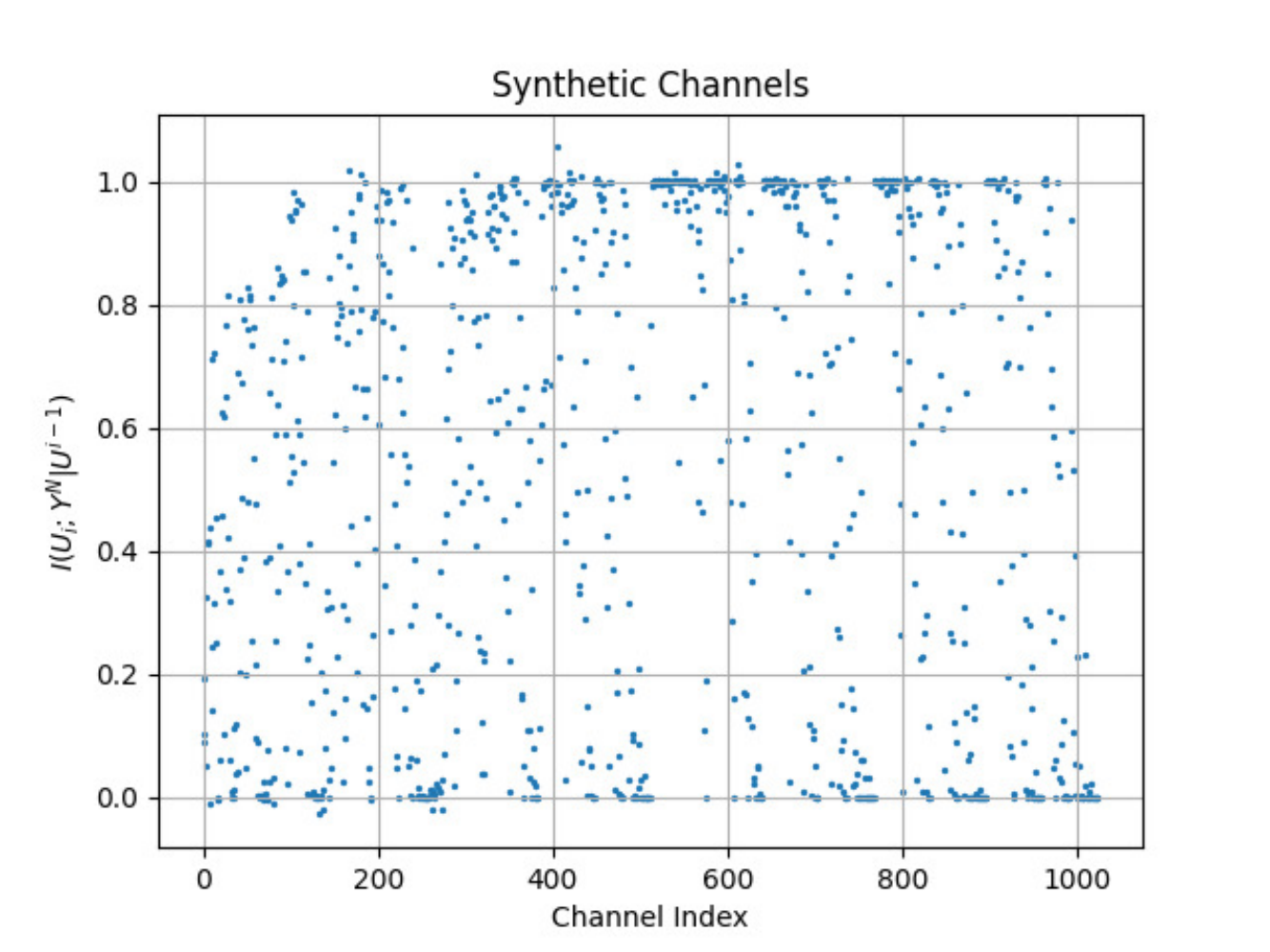}
        \caption{Polarization for the Ising channel with optimized input distribution. }
        \label{fig:polar_opt}
    \end{minipage}
    \hfill         
    \begin{minipage}{0.49\textwidth}
        \centering
        \includegraphics[width=\columnwidth, trim={0 0 0 55},clip]{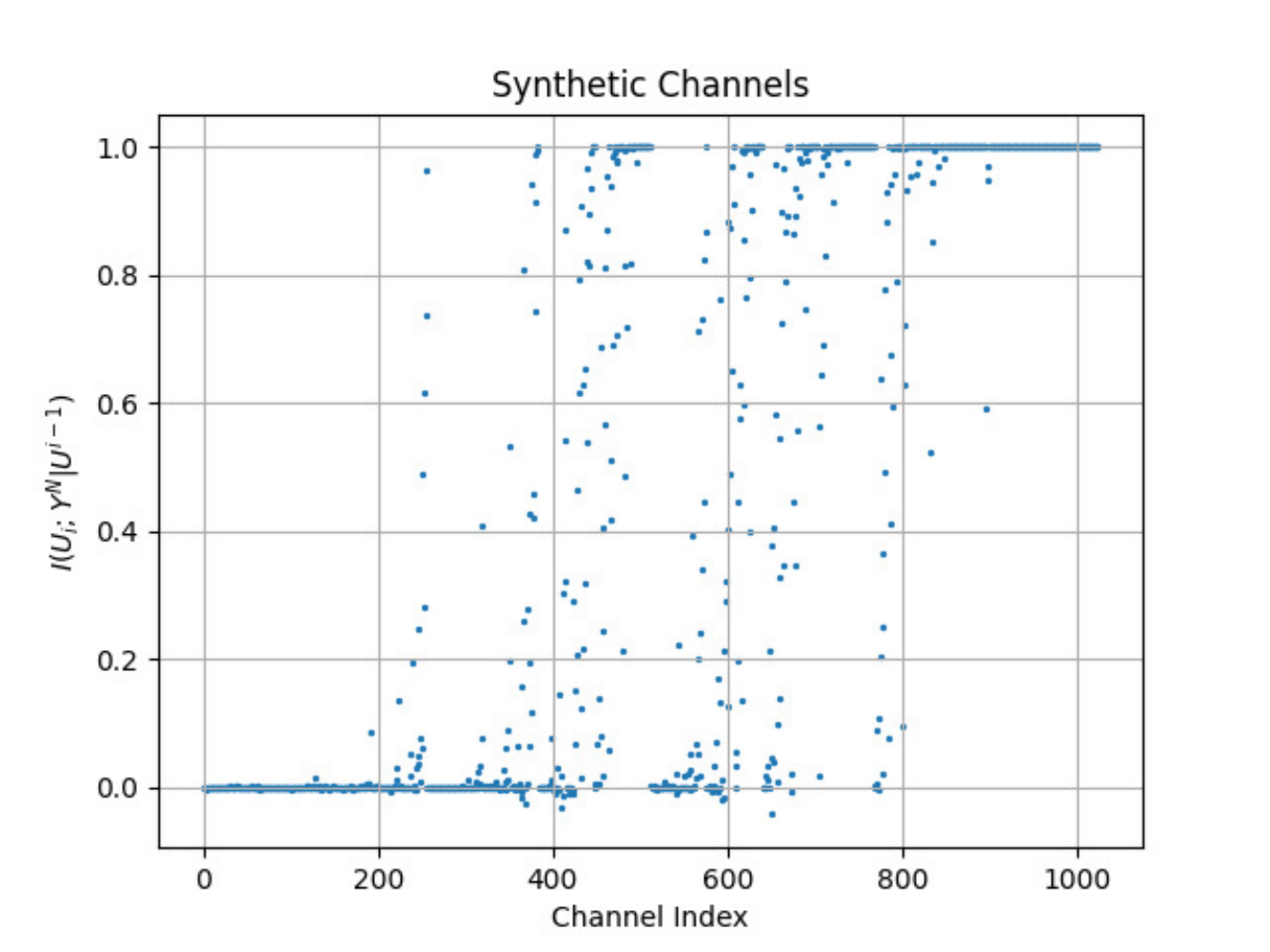}
        \caption{Polarization for the Ising channel with uniform \gls{iid} input distribution.}
        \label{fig:polar_uniform}
    \end{minipage}
\end{figure*}
\subsection{Capacity Estimation}\label{sec:applications:capacity}

\par The following experiments illustrate the \gls{mi} attained by application of Algorithm \ref{alg:npd_mi_max} on the three instances of channels. For each of these channels, the estimated \gls{mi} is recorded, as given in Equation \eqref{eqn:mi_est_step}. 
Figure \ref{fig:awgn_mi} illustrates the results obtained on the binary-input \gls{awgn} channel. 
It illustrates the evolution of the estimated \gls{mi} throughout the iterations of the algorithm. It is visible that the estimated \gls{mi} values converge to the optimal \gls{mi} for the binary-input \gls{awgn} channel. 
The figure presents results from $10$ independent simulations, each initialized with a randomly chosen value for $\psi:=P_X(1)$ within the interval $[0,1]$. Algorithm \ref{alg:npd_mi_max} is executed with $\mathsf{N_{warmup}}=1000$, and therefore, during the initial $\mathsf{N_{warmup}}$ iterations, $\psi$ is fixed. After $\mathsf{N_{warmup}}$ iterations, the algorithm starts optimizing $\psi$. Notably, after approximately 2000 iterations, the algorithm converges to the optimal value of $\f{\sI}{X;Y}$ for $P_X(1)=0.5$. 

\par Figure \ref{fig:fscs_mi} illustrates the results for the Ising and Trapdoor channels \cite{berger1990capacity}. This experiment showcases the estimated value $\frac{1}{N}\f{\what{\sI}_{\theta}}{U_\psi^N;Y^N}$ obtained upon termination of Algorithm \ref{alg:npd_mi_max} for various block lengths $N=2^n$. The solid red circles show the empirical \gls{mi} values for the optimized input distribution, $P_{X^N}^{\psi^*}$. The benchmarks for comparison in this experiment is lower and upper bounds on the capacity of the Ising channel and Trapdoor channels. The bounds are given by $0.5451\leq\mathsf{C_{Ising}}\leq 0.5482$, as derived in \cite{Ising_artyom_ISIT,huleihel2021computable}, and by $0.572\leq\mathsf{C_{Trapdoor}}\leq 0.5849$, as derived in \cite{kobayashi2003trapdoor,huleihel2021computable}. The results demonstrate that as $N$ increases, $\frac{1}{N}\f{\what{\sI}_\psi}{U^N;Y^N}$ converges toward the true value of the channel capacity. In order to emphasize the improvement of the optimized input distribution over a uniform \gls{iid} input distribution, we plotted the empirical \gls{mi} attained by the \gls{sct} decoder.
For the Ising channel, the optimized input distribution achieves \gls{mi} of $0.5415$, compared to $0.45$ for the uniform i.i.d. input. Similarly, for the Trapdoor channel, the optimized distribution attains \gls{mi} of $0.571$, whereas the uniform i.i.d. input yields $0.5$.

\par Figures \ref{fig:polar_opt} and \ref{fig:polar_uniform} illustrate the polarization of the synthetic channels for the Ising channel with optimized input distribution and uniform \gls{iid} input distribution, respectively. The figure shows that the optimized input distribution increases the average \gls{mi} of the synthetic channels on the expense of the polarization of the synthetic channels. This presents a drawback when build polar codes with the optimized input distribution, as discussed in Section \ref{sec:applications:coding}. 
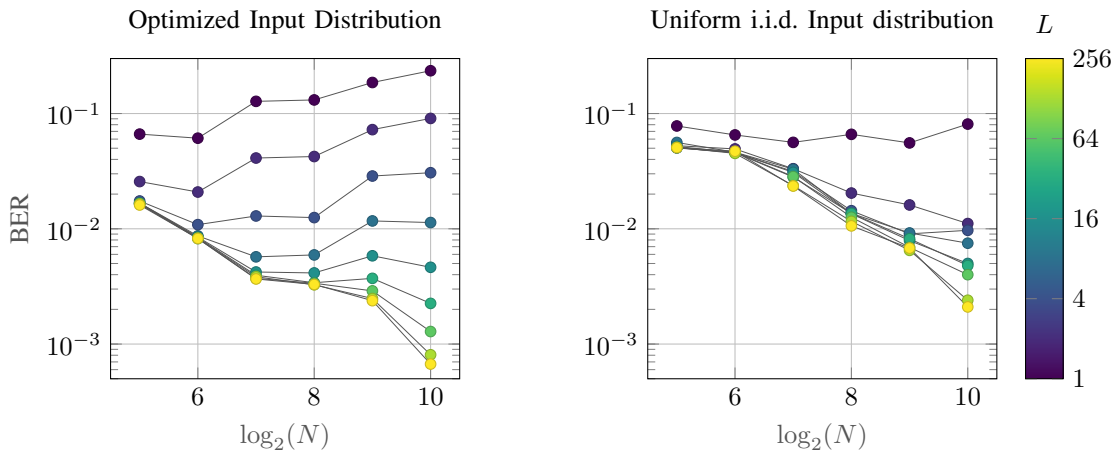
\begin{figure*}[b!]
    \centering
    \input{figures/optimized_ber_ising}
    \caption{Comparison of the BERs attained by the outcome of Algorithm \ref{alg:npd_mi_max} (left) and by the \gls{sct} decoder with uniform \gls{iid} input distribution (right) on the Ising channel. The decoding is done by \gls{scl} with values of $L=\{1,2,4,8,16,32,64,128,256\}$.}
    \label{fig:fscs_mi_ising}
\end{figure*}


\subsection{Channel Coding}\label{sec:applications:coding}
\par This section describes how the output of Algorithm~\ref{alg:npd_mi_max} is utilized to construct a polar code. At this stage, the parameters $\theta^\ast = \big\{\theta^{\ast,\mathsf{co}},\theta^{\ast,\mathsf{ch}}\big\}$ and $\psi^\ast$ are fixed and employed for code design, encoding, and decoding. During the code design phase, the frozen set $\cF$ is determined. After this step, the parameters $\psi^\ast$ are discarded, and only $\theta^\ast$ is used by the encoder and decoder. This is because the encoding uses the \gls{hy} scheme, which relies on the \gls{npd} parameterized by $\theta^{\ast,\mathsf{co}}$, which, after optimization, approximates the distribution defined by $\psi^\ast$. 

\par As illustrated in Figures~\ref{fig:polar_opt} and \ref{fig:polar_uniform}, the optimized input distribution reduces the polarization of the synthetic channels. This reduction necessitates additional coding to map the information bits onto $U_\cA$. This issue was previously identified in \cite{wang2015construction,wangFVPolarCoding2015a}, where methods such as homophonic coding and arithmetic coding were proposed to address it. In this work, we propose using an adaptive frozen set to mitigate this issue. The remainder of this section outlines the implementation of the key components for polar codes: code design, encoding, and decoding.

\subsubsection{\textbf{Code Design}}
The code design involves estimating the \gls{mi} of the synthetic channels. For this purpose, we extend Algorithm \ref{alg:npd_loss} to return $\mathbf{v}_n, \mathbf{l}_n$ instead of $\f{\cL}{x^N,y^N;\theta}$. Recall that $\mathbf{v}_n=u^N$ and $\mathbf{l}_n$ is the estimated \glspl{llr} $\{\f{L^{\theta}_{U_i|Y^N,U^{i-1}}}{y^N,u^{i-1}}\}_{i=1}^N$. Let the extension of Algorithm \ref{alg:npd_loss} be denoted by 
\begin{equation}
    [\mathbf{v}_n, \mathbf{l}_n] = \f{\wtilde{\mathsf{Alg}}_1}{x^N,y^N;\theta^\prime},
\end{equation}
where $\theta^\prime=\theta^{\ast,\mathsf{co}}$ or $\theta^\prime=\theta^{\ast,\mathsf{ch}}$.
With this notation, the code design is described by the following steps. First, we generate $\cD^{\psi^\ast}_{M,N} \sim P^{\psi^\ast}_{X^N} \otimes W_{Y^N\|X^N}$. Next, for every $x^N,y^N\in\cD^{\psi^\ast}_{M,N}$, we compute
\begin{align}
    [\mathbf{v}^\mathsf{co}_n, \mathbf{l}^\mathsf{co}_n] &= \f{\wtilde{\mathsf{Alg}}_1}{x^N,\mathbf{0}^N;\theta^{\ast,\mathsf{co}}} \\
    [\mathbf{v}^\mathsf{ch}_n, \mathbf{l}^\mathsf{ch}_n] &= \f{\wtilde{\mathsf{Alg}}_1}{x^N,y^N;\theta^{\ast,\mathsf{ch}}}.
\end{align}
Then, the empirical \gls{mi} of the synthetic channels is computed for all $i\in[1:N]$
\begin{align}
    \f{\what{\sI}_{\theta^\ast}}{U_i;Y^N|U^{i-1}} &= \\ 
    &\hspace{-2.65cm}\frac{1}{M}\hspace{-0.7cm}\sum_{x^N,y^N\in\cD^{\psi^\ast}_{M,N}}\hspace{-0.65cm} \Big[ -v^\mathsf{co}_{n,i} \log \sigma(l^\mathsf{co}_{n,i}) - (1-v^\mathsf{co}_{n,i}) \log (1 - \sigma(l^\mathsf{co}_{n,i}))\Big] \nn\\ 
    &\hspace{-2.85cm}-\hspace{-0.1cm} \Big[\hspace{-0.1cm} -v^\mathsf{ch}_{n,i} \log \sigma(l^\mathsf{ch}_{n,i}) - (1-v^\mathsf{ch}_{n,i}) \log (1 - \sigma(l^\mathsf{ch}_{n,i}))\Big].
\end{align}
Given the design rate $R\in[0,1]$, we set the information set $\cA$ to include the indices with the highest $k=\lfloor RN\rfloor$ values of $\f{\what{\sI}_{\theta^\ast}}{U_i;Y^N|U^{i-1}}$ and the frozen set, implied by $\cF=[1:N]\setminus \cA$.

\subsubsection{\textbf{Encoder with Fixed Frozen Set}}
Let $b^k$ be uniform \gls{iid} information bits and let $\cF\subset [1:N]$ be the frozen set. In order to encode $b^k$ into $x^N$, the encoder uses the recursive formulas of the \gls{sc} decoding of the \gls{npd} with the parameters $\theta^{\ast,\mathsf{co}}$. That is, the encoder uses the constant channel embedding to compute the \glspl{llr} $\{\f{L^{\theta^{\ast,\mathsf{co}}}_{U_i|U^{i-1}}}{u^{i-1}}\}_{i=1}^N$ sequentially. The only difference is what happens when the decoder reaches the decision stage. Here, the encoder assigns an information bit or decides on the value of the frozen bits. Thus, upon reaching the $i$\textsuperscript{th} decision
\begin{align}
    u_i = \begin{cases}
        \f{\mathsf{sgn}}{\f{L^{\theta^{\ast,\mathsf{co}}}_{U_i|U^{i-1}}}{u^{i-1}}}  &i\in\cF \\
        b_{j(i)}                   &i\in\cA,
    \end{cases}
\end{align}
where $j(i) = |\{l\in\mathcal{A}: l\le i\}|$ and $\mathsf{sgn}$ is the sign function.

\subsubsection{\textbf{Encoder with Adaptive Frozen Set}}
Let $b_1, b_2, \dots$ be information bits the encoder needs to transmit across multiple blocks. An encoder with an adaptive frozen set operates similarly to one with a fixed frozen set, but it allows for the frozen positions to vary from block to block. Specifically, the encoder decides whether to place an information bit or a shaping bit in each position $i\in\cA$ by inspecting the \gls{llr} $L^{\theta^{\ast,\mathsf{co}}}_{U_i|U^{i-1}}(u^{i-1})$. If the \gls{llr} is close to $0$, it indicates that $P^{\theta^{\ast,\mathsf{co}}}_{U_i|U^{i-1}}(\cdot|u^{i-1})$ is approximately uniform, and an information bit can be inserted without distorting the codeword distribution from the optimized one. Conversely, if the \gls{llr} is large in magnitude, the conditional distribution is biased, and inserting an information bit would shift the codeword distribution. In such cases, the encoder inserts a shaping bit sampled according to $P_{U_i|U^{i-1}}$. We define the adaptive frozen by
\begin{equation}
    \cF_d = \{i\in[1:N]: i\in\cF \text{ or } |L^{\theta^{\ast,\mathsf{co}}}_{U_i|U^{i-1}}(u^{i-1})|> t\},
\end{equation}
and accordingly, $\cA_d = [N] \setminus \cF_d$, where $t>0$ is a predetermined threshold. The code rate in this setting is computed by $\bE[|\cA_d|]/N$.
\subsubsection{\textbf{Decoder}}
The decoder observes $y^N$ and uses the parameters $\theta^{\ast,\mathsf{co}}, \theta^{\ast,\mathsf{ch}}$ to decode $u^N$. This amounts to apply \gls{sc} decoding of the \gls{npd} with the parameters $\theta^{\ast,\mathsf{co}}, \theta^{\ast,\mathsf{ch}}$ simultaneously. That is, the decoder uses both embeddings to compute  the \glspl{llr} $\{\f{L^{\theta^{\ast,\mathsf{co}}}_{U_i|U^{i-1}}}{u^{i-1}}\}_{i=1}^N$ and $\{\f{L^{\theta^{\ast,\mathsf{ch}}}_{U_i|Y^N,U^{i-1}}}{y^N,u^{i-1}}\}_{i=1}^N$. Upon reaching the decision stage, the decoder makes it decision according to the following rule
\begin{align}
    \hat{u}_i = \begin{cases}
        \f{\mathsf{sgn}}{\f{L^{\theta^{\ast,\mathsf{co}}}_{U_i|U^{i-1}}}{\hat{u}^{i-1}}}  &i\in\cF \\
        \f{\mathsf{sgn}}{\f{L^{\theta^{\ast,\mathsf{ch}}}_{U_i|Y^N,U^{i-1}}}{y^N,\hat{u}^{i-1}}}                           &i\in\cA.
    \end{cases}
\end{align}
In the adaptive frozen set setting, since the decoder computes also $L_{U_i|U^{i-1}}(u^{i-1})$, it can also determine the adaptive frozen set $\cF_d$. A block error is declared if $\hat{u}_i \neq u_i$ for any $i\in\cA_d$ or the decoder fails to discover the information set $\cA_d$.
\subsubsection{\textbf{Experiments}}
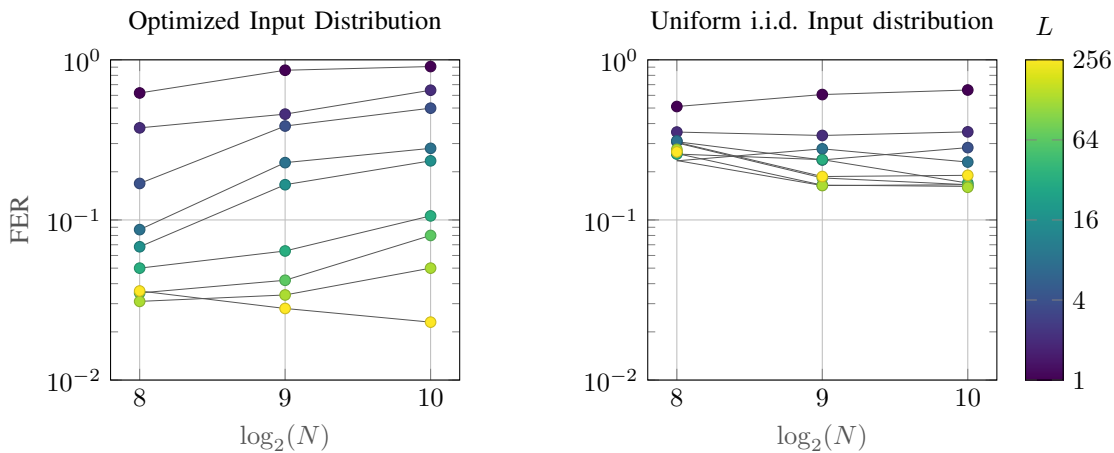
\begin{figure*}[b!]
    \centering
    \input{figures/optimized_fer_ising.tex}
    \caption{Comparison of the FERs attained by the outcome of Algorithm \ref{alg:npd_mi_max} (left) and by the \gls{sct} decoder with uniform \gls{iid} input distribution (right) on the Ising channel. The optimized input distribution uses adaptive frozen set with $t=1$. The decoding is done by \gls{scl} with values of $L=\{1,2,4,8,16,32,64,128,256\}$}
    \label{fig:fer_adaptive_ising}
\end{figure*}


\par The following experiments illustrate the incurred \glspl{ber} of \gls{scl} decoding for various lists size $L=\{1,2,4,8,16,32,64,128,256\}$ ($L=1$ is \gls{sc} decoding) on the Ising channel. The experiments compare the attained \glspl{ber} by a uniform \gls{iid} input distribution and \gls{sct} list decoder, which provides a theoretical benchmark, and the \glspl{ber} attained by \gls{scl} decoding of the \gls{npd} with parameter $\theta^\ast$, as obtained by Algorithm \ref{alg:npd_mi_max}. 

\par Figure~\ref{fig:fscs_mi_ising} illustrates the decoding errors for the Ising channel for the case where the bits on the information set distribute according to $P^{\theta^{\ast,\mathsf{co}}}_{U_i|U^{i-1}}(\cdot|u^{i-1})$. The goal of this experiment is to examine the benefit of optimizing the input distribution when there is no rate loss when encoding information bits on $\cA$. For the optimized input distribution, the size of the information was set to $\lfloor 0.4N \rfloor$. Since the optimized input distribution is not necessarily uniform, the sum of conditional entropies $\sum_{i\in\cA}\sH (P^{\theta^{\ast,\mathsf{co}}}_{U_i|U^{i-1}})$ may be strictly less than $0.4N$. Hence, to make a fair comparison between uniform and optimized input distribution, we set the code rate of uniform and \gls{iid} input distribution to be $\frac{1}{N}\sum_{i\in\cA}\sH (P^{\theta^{\ast,\mathsf{co}}}_{U_i|U^{i-1}})$. Figure~\ref{fig:fscs_mi_ising} shows the optimized input distribution attains lower \glspl{ber} for code rate of 0.4 for the optimized input distribution and approximately $0.37$ for the \gls{iid} inputs.

\par Figure~\ref{fig:fer_adaptive_ising} shows the \glspl{fer} for the Ising channel when using adaptive frozen sets. In this experiments the uniform input distribution uses code rate of $0.4$ and the optimized input distribution uses code rate of $0.435$ and threshold $1$, which yielded an expected rate of $0.4$. Under this setting we observe better performance of the optimized input distribution than the uniform and \gls{iid} input distribution.

\par Since $R$ is close to the \gls{mi}, we observe non-monotonic behavior in the \gls{ber} as a function of block length, particularly for small list sizes. In these cases, increasing the block length leads to performance degradation due to the limitations of the \gls{sc} decoder at moderate block lengths. This phenomenon is observed not only for the \gls{npd}, but also for the \gls{sct} decoder, which leverages the channel model to compute exact posteriors of the synthetic channels. This further supports the interpretation that the degradation arises from inherent limitations of the \gls{sc} decoding principle rather than the specific decoder implementation.

\section{Conclusions and Future Work}\label{sec:conc}
\par In this paper, we proposed a novel method to optimize the input distribution of polar codes using \glspl{npd}. Our approach addresses the dual challenges of maximizing code rates over input distributions and providing a practical coding scheme within the framework of polar codes. This is particularly valuable for scenarios where the channel model is unknown, treating the channel as a black-box that generates output samples from input samples.

\par The methodology involves a two-phase process: a training phase and an inference phase. During the training phase, we alternate between estimating \gls{mi} using \glspl{npd} and optimizing the input distribution parameters. In the inference phase, the optimized model is used to construct polar codes, incorporating the \gls{hy} scheme and list decoding to enhance performance.

\par Our experimental results on memoryless channels and \glspl{fsc} demonstrated the effectiveness of our approach. We observed significant improvements in \gls{mi} and \glspl{ber} compared to uniform and \gls{iid} input distributions, validating our method for block lengths up to $1024$. The scalability of our approach suggests its potential applicability to real-world communication systems, bridging the gap between theoretical capacity estimation and practical coding performance.

\par Future work will focus on  mitigating the reduced polarization caused by optimized inputs by developing techniques that improve code rates without compromising the polarization of synthetic channels.

\par Overall, our work contributes to advancing the understanding and application of \glspl{npd} in channel coding, providing a robust framework for high-performance communication systems. Future research directions include exploring further optimizations and extensions of our methodology to other types of channels and coding schemes.

\bibliography{ref,ref-new}
\end{document}

%% file: figures/awgn_mi.tex
\begin{tikzpicture}

\definecolor{chocolate2267451}{RGB}{226,74,51}
\definecolor{dimgray85}{RGB}{85,85,85}
\definecolor{gainsboro229}{RGB}{229,229,229}

\begin{axis}[
title=AWGN,
log basis y={10},
x grid style={gainsboro229},
xlabel=\textcolor{dimgray85}{Iterations},
xmajorgrids,
xtick style={color=dimgray85},
y grid style={gainsboro229},
ylabel={$\displaystyle \frac{1}{N}\what{\mathsf{I}}_{\theta}\left(U_\psi^N;Y^N\right)$},
ymajorgrids,
xmode=log,
ytick style={color=dimgray85},
legend style={at={(0.05,0.05)},anchor=south west, nodes={scale=0.85, transform shape} },
]

    \foreach \i in {1,2,...,8} {
        \addplot[smooth, line width=0.05pt, violet] table [x=iter, y=c\i, col sep=comma] {figures/awgn_mi.csv};
        }
        \draw [black, dashed, line width=0.75pt] (axis cs:1000,0) -- (axis cs:1000,1);
        \draw [black, thick] (axis cs:1,0.6231) -- (axis cs:20000,0.6231);
        \node at (axis cs:250,0.55) {Estimation};
        \node at (axis cs:8000,0.55) {Estimation};
        \node at (axis cs:8000,0.5) {+};
        \node at (axis cs:8000,0.45) {Improvement};

\end{axis}

\end{tikzpicture}

%% file: figures/nsc-optimize.tex
\begin{tikzpicture}
\definecolor{darkblue}{rgb}{0,0,0.55}
\definecolor{darkgreen}{rgb}{0,0.39,0}
\definecolor{maroon}{rgb}{0.5,0,0}
\definecolor{plum}{rgb}{0.56,0.27,0.52}
\definecolor{darkorange}{rgb}{0.8,0.33,0} 
\definecolor{dimgray85}{RGB}{85,85,85}
\definecolor{gainsboro229}{RGB}{229,229,229}


\begin{axis}[
    name=plot1,
    title=Ising,
    log basis y={10},
    x grid style={gainsboro229},
    xlabel=\textcolor{dimgray85}{$\displaystyle \log_2(N)$},
    xmajorgrids,
    ymin=0.41, ymax=0.555,
    xtick style={color=dimgray85},
    y grid style={gainsboro229},
    ylabel=\textcolor{dimgray85}{$\displaystyle \frac{1}{N}\what{\mathsf{I}}_\theta\left(U_\psi^N;Y^N\right)$},
    ymajorgrids,
    domain=5:10,
    xtick={5,6,7,8,9,10},
    ytick style={color=dimgray85},
    legend style={at={(0.95,0.05)},anchor=south east, nodes={scale=0.85, transform shape} }
    ]
\addplot [semithick, dashed, maroon]
table{
5  0.442736953
6  0.44658398628234863
7  0.44888442754745483
8  0.44996562600135803
9  0.450315386
10 0.45058703422546387
};
\addplot [only marks, maroon]
table{
5  0.5323
6  0.5385
7  0.5397
8  0.5403
9  0.5408
10 0.5415
};
\addplot[semithick, black,
        mark=triangle, 
        mark options={rotate=180, color=black, scale=1.0}, mark repeat=4] {0.5451};
\addplot[semithick, black,
        mark=triangle, 
        mark options={rotate=0, color=black, scale=1.0}, mark repeat=4] {0.5482};
\end{axis}

\begin{axis}[
    name=plot2,
    at={(plot1.east)},
    anchor=west,
    xshift=2.5cm, 
    title=Trapdoor,
    log basis y={10},
    x grid style={gainsboro229},
    xlabel=\textcolor{dimgray85}{$\displaystyle \log_2(N)$},
    xmajorgrids,
    ymin=0.42, ymax=0.59,
    xtick style={color=dimgray85},
    y grid style={gainsboro229},
    ymajorgrids,
    xtick={5,6,7,8,9,10},
    domain=5:10,
    ytick style={color=dimgray85},
    legend style={at={(-.52,0.25)},anchor=south west, nodes={scale=0.85, transform shape} }
    ]
\addplot [semithick, dashed, maroon]
table{
5  0.46883898973464966
6  0.4843657910823822
7  0.49204933643341064
8  0.49621760845184326
9  0.49791476130485535
10 0.4990154504776001
};
\addlegendentry{i.i.d.}
\addplot [only marks, maroon]
table{
5  0.5381
6  0.5537
7  0.5627
8  0.5674
9  0.5699
10 0.5710
};
\addlegendentry{ $P^{\psi^\ast}_{X^N}$}
\addplot[semithick, black,
        mark=triangle, 
        mark options={rotate=180, draw=black, scale=1.0}, mark repeat=4] {0.572};
\addlegendentry{LB}
\addplot[semithick, black,
        mark=triangle, 
        mark options={rotate=0, color=black, scale=1.0}, mark repeat=4] {0.5849};
\addlegendentry{UB}
\end{axis}

\end{tikzpicture}

%% file: figures/optimized_ber_ising.tex
\begin{tikzpicture}
\definecolor{darkblue}{rgb}{0,0,0.55}
\definecolor{darkgreen}{rgb}{0,0.39,0}
\definecolor{maroon}{rgb}{0.5,0,0}
\definecolor{plum}{rgb}{0.56,0.27,0.52}
\definecolor{darkorange}{rgb}{0.8,0.33,0} 
\definecolor{dimgray85}{RGB}{85,85,85}
\definecolor{gainsboro229}{RGB}{229,229,229}

\begin{axis}[
    name=plot1,
    title={Optimized Input Distribution},
    xlabel={\textcolor{dimgray85}{$\displaystyle \log_2(N)$}},
    ylabel={\textcolor{dimgray85}{BER}},
    ymode=log,
    ymin=5e-4, ymax=0.3,
    xmajorgrids,
    ymajorgrids,
    colormap/viridis,
  ]
    \addplot [scatter,
      only marks,
      scatter src=explicit,
    ]
    table[meta expr=log10(\thisrow{meta})/log10(2)]{
n	ber	meta

5	0.06629935	1
5	0.025696318692144664	2
5	0.017428785806846644	4
5	0.016709145630734494	8
5	0.016692487289761377	16
5	0.016605031018790933	32
5	0.016449275560777583	64
5	0.016458437647791518	128
5	0.016085291	256
6	0.061116642	1
6	0.020843178262298312	2
6	0.010864567562066454	4
6	0.008676062	8
6	0.008474563	16
6	0.008370215	32
6	0.008215492	64
6	0.008290255	128
6	0.008245077	256
7	0.12754152651340037	1
7	0.041108466400925456	2
7	0.012919422964389058	4
7	0.005710674	8
7	0.004218675	16
7	0.003954101	32
7	0.003712066	64
7	0.003839257	128
7	0.003671694	256
8	0.13120097094537994	1
8	0.042359018	2
8	0.012508353983610124	4
8	0.005933798	8
8	0.004140381	16
8	0.003405944	32
8	0.003370374	64
8	0.003267386	128
8	0.00329012	256
9	0.1859047440439323	1
9	0.072498114	2
9	0.028696583345581752	4
9	0.011699738529972036	8
9	0.005831986	16
9	0.003715201	32
9	0.002892965	64
9	0.002483954	128
9	0.002379154	256
10	0.23510296203503464	1
10	0.090695972	2
10	0.030658045058436077	4
10	0.011330618318891724	8
10	0.0046336	16
10	0.002252786	32
10	0.00128437	64
10	0.000807591	128
10	0.000670569	256
};

\addplot[dimgray85, line width=0.1pt] coordinates {(5,0.06629935) (6,0.061116642) (7,0.12754152651340037) (8,0.13120097094537994) (9,0.1859047440439323) (10,0.23510296203503464)};
\addplot[dimgray85, line width=0.1pt] coordinates { (5,0.025696318692144664) (6,0.020843178262298312) (7,0.041108466400925456) (8,0.042359018) (9,0.072498114) (10,0.090695972)};
\addplot[dimgray85, line width=0.1pt] coordinates { (5,0.017428785806846644) (6,0.010864567562066454) (7,0.012919422964389058) (8,0.012508353983610124) (9,0.028696583345581752) (10,0.030658045058436077)};
\addplot[dimgray85, line width=0.1pt] coordinates {(5,0.016709145630734494) (6,0.008676062) (7,0.005710674) (8,0.005933798) (9,0.011699738529972036) (10,0.011330618318891724)};
\addplot[dimgray85, line width=0.1pt] coordinates {(5,0.016692487289761377) (6,0.008474563) (7,0.004218675) (8,0.004140381) (9,0.005831986) (10,0.0046336)};
\addplot[dimgray85, line width=0.1pt] coordinates {(5,0.016605031018790933) (6,0.008370215) (7,0.003954101) (8,0.003405944) (9,0.003715201) (10,0.002252786)};
\addplot[dimgray85, line width=0.1pt] coordinates { (5,0.016449275560777583) (6,0.008215492) (7,0.003712066) (8,0.003370374) (9,0.002892965) (10,0.00128437)};
\addplot[dimgray85, line width=0.1pt] coordinates { (5,0.016458437647791518) (6,0.008290255) (7,0.003839257) (8,0.003267386) (9,0.002483954) (10,0.000807591)};
\addplot[dimgray85, line width=0.1pt] coordinates {(5,0.016085291) (6,0.008245077) (7,0.003671694) (8,0.00329012) (9,0.002379154) (10,0.000670569)};

\end{axis}

\begin{axis}[
    name=plot2,
    at={(plot1.east)},
    anchor=west,
    xshift=2.5cm, 
    title={Uniform i.i.d. Input distribution},
    xlabel={\textcolor{dimgray85}{$\displaystyle \log_2(N)$}},
    ymode=log,
    ymin=5e-4, ymax=0.3,
    xmajorgrids,
    ymajorgrids,
    colorbar,
    colormap/viridis,
    colorbar style={
      title={$L$},
      ytick={0, 2, 4, 6, 8},
      yticklabels={$1$, $4$, $16$, $64$, $256$},
    },
  ]
    \addplot [scatter,
      only marks,
      scatter src=explicit,
    ]
    table[meta expr=log10(\thisrow{meta})/log10(2)]{
n	ber	meta
5	0.0780	1
5	0.0526	2
5	0.0503	4
5	0.0561	8
5	0.0518	16
5	0.0505	32
5	0.0508	64
5	0.0509	128
5	0.0506	256
6	0.0652	1
6	0.0493	2
6	0.0459	4
6	0.0462	8
6	0.0459	16
6	0.0458	32
6	0.0462	64
6	0.0451	128
6	0.0471	256 
7	0.0562	1
7	0.0332	2
7	0.03145	4
7	0.03291	8
7	0.03087	16
7	0.0281	32
7	0.0284	64
7	0.0237	128
7	0.0235	256
8	0.066	1
8	0.020419	2
8	0.014355	4
8	0.0136	8
8	0.0135	16
8	0.0136	32
8	0.0126	64
8	0.0116	128
8	0.0106	256
9	0.0556	1
9	0.0161	2
9	0.00904	4
9	0.0092	8
9	0.0079	16
9	0.0082	32
9	0.0069	64
9	0.0065	128
9	0.0068	256
10	0.081	1
10	0.0111	2
10	0.0097	4
10	0.0075	8
10	0.005	16
10	0.0048	32
10	0.0040	64
10	0.0024	128
10	0.0021	256
};

\addplot[dimgray85, line width=0.1pt] coordinates { (5,0.0780) (6,0.0652) (7,0.0562) (8,0.066) (9,0.0556) (10,0.081)};
\addplot[dimgray85, line width=0.1pt] coordinates {(5,0.0526) (6,0.0493) (7,0.0332) (8,0.020419) (9,0.0161) (10,0.0111)};
\addplot[dimgray85, line width=0.1pt] coordinates { (5,0.0503) (6,0.0459) (7,0.03145) (8,0.014355) (9,0.00904) (10,0.0097)};
\addplot[dimgray85, line width=0.1pt] coordinates { (5,0.0561) (6,0.0462) (7,0.03291) (8,0.0136) (9,0.0092) (10,0.0075)};
\addplot[dimgray85, line width=0.1pt] coordinates {(5,0.0518) (6,0.0459) (7,0.03087) (8,0.0135) (9,0.0079) (10,0.005)};
\addplot[dimgray85, line width=0.1pt] coordinates { (5,0.0505) (6,0.0458) (7,0.0281) (8,0.0136) (9,0.0082) (10,0.0048)};
\addplot[dimgray85, line width=0.1pt] coordinates {(5,0.0508) (6,0.0462) (7,0.0284) (8,0.0126) (9,0.0069) (10,0.0040)};
\addplot[dimgray85, line width=0.1pt] coordinates { (5,0.0509) (6,0.0451) (7,0.0237) (8,0.0116) (9,0.0065) (10,0.0024)};
\addplot[dimgray85, line width=0.1pt] coordinates { (5,0.0506) (6,0.0471) (7,0.0235) (8,0.0106) (9,0.0068) (10,0.0021)};
\end{axis}

\end{tikzpicture}

%% file: figures/optimized_fer_ising.tex
\begin{tikzpicture}
\definecolor{darkblue}{rgb}{0,0,0.55}
\definecolor{darkgreen}{rgb}{0,0.39,0}
\definecolor{maroon}{rgb}{0.5,0,0}
\definecolor{plum}{rgb}{0.56,0.27,0.52}
\definecolor{darkorange}{rgb}{0.8,0.33,0} 
\definecolor{dimgray85}{RGB}{85,85,85}
\definecolor{gainsboro229}{RGB}{229,229,229}

\begin{axis}[
    name=plot1,
    title={Optimized Input Distribution},
    xlabel={\textcolor{dimgray85}{$\displaystyle \log_2(N)$}},
    ylabel={\textcolor{dimgray85}{FER}},
    ymode=log,
    ymin=1e-2, ymax=1.0,
        xmajorgrids,
    ymajorgrids,
    colormap/viridis,
  ]
    \addplot [scatter,
      only marks,
      scatter src=explicit,
    ]
    table[meta expr=log10(\thisrow{meta})/log10(2)]{
n	fer	meta
8	0.621	1
8	0.376	2
8	0.169	4
8	0.087	8
8	0.068	16
8	0.05	32
8	0.035	64
8	0.031	128
8	0.036	256
9	0.861	1
9	0.458	2
9	0.386	4
9	0.228	8
9	0.166	16
9	0.064	32
9	0.042	64
9	0.034	128
9	0.028	256
10	0.908	1
10	0.646	2
10	0.499	4
10	0.28	8
10	0.234	16
10	0.106	32
10	0.08	64
10	0.05	128
10	0.023	256
};
    Adding lines between points with the same meta value
    \addplot[dimgray85, line width=0.1pt] coordinates {(8,0.621) (9,0.861) (10,0.908)};
    \addplot[dimgray85, line width=0.1pt] coordinates {(8,0.376) (9,0.458) (10,0.646)};
    \addplot[dimgray85, line width=0.1pt] coordinates {(8,0.169) (9,0.386) (10,0.499)};
    \addplot[dimgray85, line width=0.1pt] coordinates { (8,0.087) (9,0.228) (10,0.28)};
    \addplot[dimgray85, line width=0.1pt] coordinates { (8,0.068) (9,0.166) (10,0.234)};
    \addplot[dimgray85, line width=0.1pt] coordinates {(8,0.05) (9,0.064) (10,0.106)};
    \addplot[dimgray85, line width=0.1pt] coordinates {(8,0.035) (9,0.042) (10,0.08)};
    \addplot[dimgray85, line width=0.1pt] coordinates {(8,0.031) (9,0.034) (10,0.05)};
    \addplot[dimgray85, line width=0.1pt] coordinates {(8,0.036) (9,0.028) (10,0.023)};

\end{axis}

\begin{axis}[
    name=plot2,
    at={(plot1.east)},
    anchor=west,
    xshift=2.5cm, 
    title={Uniform i.i.d. Input distribution},
    xlabel={\textcolor{dimgray85}{$\displaystyle \log_2(N)$}},
    xmajorgrids,
    ymajorgrids,
    ymode=log,
    ymin=1e-2, ymax=1.0,
    colorbar,
    colormap/viridis,
    colorbar style={
      title={$L$},
      ytick={0, 2, 4, 6, 8},
      yticklabels={$1$, $4$, $16$, $64$, $256$},
    },
  ]
    \addplot [scatter,
      only marks,
      scatter src=explicit,
    ]
    table[meta expr=log10(\thisrow{meta})/log10(2)]{
n	ber	meta
8	0.511	1
8	0.354	2
8	0.308	4
8	0.312	8
8	0.26	16
8	0.258	32
8	0.2676	64
8	0.276	128
8	0.266	256
9	0.607	1
9	0.337	2
9	0.237	4
9	0.278	8
9	0.165	16
9	0.238	32
9	0.183	64
9	0.164	128
9	0.1867	256
10	0.648	1
10	0.355	2
10	0.283	4
10	0.23	8
10	0.162	16
10	0.17	32
10	0.166	64
10	0.16	128
10	0.19	256
};

    \addplot[dimgray85, line width=0.1pt] coordinates { (8,0.511) (9,0.607) (10,0.648)};
    \addplot[dimgray85, line width=0.1pt] coordinates { (8,0.354) (9,0.337) (10,0.355)};
    \addplot[dimgray85, line width=0.1pt] coordinates { (8,0.308) (9,0.237) (10,0.283)};
    \addplot[dimgray85, line width=0.1pt] coordinates { (8,0.234) (9,0.278) (10,0.23)};
    \addplot[dimgray85, line width=0.1pt] coordinates { (8,0.26) (9,0.165) (10,0.162)};
    \addplot[dimgray85, line width=0.1pt] coordinates {(8,0.258) (9,0.238) (10,0.17)};
    \addplot[dimgray85, line width=0.1pt] coordinates {(8,0.303) (9,0.183) (10,0.166)};
    \addplot[dimgray85, line width=0.1pt] coordinates {(8,0.235) (9,0.164) (10,0.166)};
    \addplot[dimgray85, line width=0.1pt] coordinates { (8,0.306) (9,0.1867) (10,0.19)};
\end{axis}

\end{tikzpicture}